\documentclass[10pt,journal,compsoc,review]{IEEEtran}

\usepackage{booktabs} 
\usepackage{enumitem}
\usepackage{xcolor}
\usepackage{ifthen}
\usepackage{amssymb}
\usepackage{multirow}
\usepackage{multicol}
\usepackage{amsthm}
\usepackage{url}
\usepackage{pgfplots}
\usepackage{hyperref}
\usepackage{balance}
\usepackage{makecell}

\usepackage{amsmath}
\usepackage{commath}
\usepackage{listings}
\definecolor{codegreen}{rgb}{0,0.6,0}
\definecolor{codegray}{rgb}{0.5,0.5,0.5}
\definecolor{codepurple}{rgb}{0.58,0,0.82}
\definecolor{backcolour}{rgb}{0.95,0.95,0.92}
\usepackage[skip=0cm,list=true,labelfont=it]{subcaption}

\usepackage[most]{tcolorbox}

\lstdefinestyle{mystyle}{
    backgroundcolor=\color{backcolour},
    commentstyle=\color{codegreen},
    keywordstyle=\color{magenta},
    numberstyle=\tiny\color{codegray},
    stringstyle=\color{codepurple},
    basicstyle=\footnotesize,
    breakatwhitespace=false,
    breaklines=true,
    captionpos=b,
    keepspaces=true,
    numbers=left,
    numbersep=5pt,
    showspaces=false,
    showstringspaces=false,
    showtabs=false,
    tabsize=2
}

\theoremstyle{definition}
\newtheorem{finding}{Finding}

\newcommand{\find}[1]{\begin{tcolorbox}[left=2pt,right=2pt,top=2pt,bottom=2pt]
\begin{finding}
#1
\end{finding}
\end{tcolorbox}
}

\newboolean{showcomments}
\setboolean{showcomments}{true}
\ifthenelse{\boolean{showcomments}}
 { \newcommand{\mynote}[2]{
      \fbox{\bfseries\sffamily\scriptsize#1}
        {\small$\blacktriangleright$\textsf{\emph{#2}}$\blacktriangleleft$}}}
        { \newcommand{\mynote}[2]{}}

\newcommand{\toolname}{{\tt D\&C}\xspace}

\usepackage[ruled]{algorithm2e} 

\SetAlFnt{\small}
\SetAlCapFnt{\small}
\SetAlCapNameFnt{\small}
\SetAlCapHSkip{0pt}
\IncMargin{-\parindent}

\begin{document}
\title{D\&C: A Divide-and-Conquer Approach to \\IR-based Bug Localization}

\author{Anil~Koyuncu, 
		Tegawend\'e~F.~Bissyand\'e, 
		Dongsun~Kim, \\  
		Kui~Liu, 
		Jacques~Klein, 
		 Martin~Monperrus 
		and Yves~Le~Traon

\IEEEcompsocitemizethanks{\IEEEcompsocthanksitem A. Koyuncu, T.F. Bissyand\'e, D. Kim,
K. Liu, J. Klein and Y. Le~Traon
 are with SnT, University of Luxembourg, Luxembourg.
\IEEEcompsocthanksitem M. Monperrus is with KTH Royal Institute of Technology, Sweden.}
}
%

\markboth{IEEE TRANSACTIONS ON SOFTWARE ENGINEERING}%
{Koyuncu \MakeLowercase{\textit{et al.}}: \toolname: A boosting Approach}

\IEEEtitleabstractindextext{%
\begin{abstract}
Many automated tasks in software maintenance rely on information retrieval (IR) techniques to identify specific information within unstructured data.
Bug localization is such a typical task, where text in a bug report is analyzed
to identify file locations in the source code that can be associated to the reported bug.
Unfortunately, despite the promising results reported in
the literature, the performance offered by IR-based bug localization tools is still not significant for large adoption. We argue that one reason could be the attempt by the community to build a ``one-size-fits-all'' approach for bug localization, without fully addressing the differences of available information that may exist among the bug reports and across the project source code files.

In this paper, we first extensively study the performance of state-of-the-art bug localization tools, specifically focusing on investigating the query formulation (i.e., which bug report features should be compared against which features of source code files) and its importance with respect to the localization performance. Building on insights from this study, we propose a new learning approach where multiple classifier models are trained on clear-cut sets of bug-location pairs. Concretely, we apply a gradient boosting supervised learning approach to various sets of bug reports whose localizations appear to be successful with specific types of features. The training scenario builds on our findings that the various state-of-the-art localization tools (hence the associated similarity features that they leverage) can be highly performant for specific sets of bug reports. We implement \toolname, a multi-classifier approach, which computes appropriate weights that should be assigned to the similarity measurements between pairs of information token types (the bug report and source code). Experimental results on large and up-to-date datasets reveal that \toolname outperforms state-of-the-art tools. On average, the validation experiments yield an MAP score of 0.52, and an MRR score of 0.63 with a curated dataset. Comparison against the state-of-the-art shows that \toolname provides a substantial performance improvement of MAP and MRR over all tools: MAP is improved by between 4 and up to 10 percentage points, while MRR is improved by between 1 and up to 12. Finally, we note that \toolname is stable in its localization performance: around 50\% of bugs can be located at Top1, 77\% at Top5 and 85\% at Top10.

\end{abstract}


}

%
%

%
%


\maketitle



\IEEEraisesectionheading{\section{Introduction}
\label{intro}}
\IEEEPARstart{B}{ug} tracking is now commonplace in software development ecosystems. Dedicated systems are set up by development teams in large-scale systems (e.g., Linux) and smaller software projects alike~\cite{bissyande2013got}. Bug tracking systems (such as Bugzilla\footnote{\url{https://www.bugzilla.org}} and Jira\footnote{\url{https://jira.atlassian.com}}) implement a communication channel between developers and software users, and are used by developers themselves to keep track of the bugs that they encounter. A typical bug report is a natural language description of a problem that a user has encountered while interacting with the software product, including more or less details on how to reproduce the bug. This report may further include other structural information such as the stack trace that was produced during a crashed execution. To fix the reported bug, developers must analyze it and eventually locate the relevant buggy code.

Automated bug localization aims at reducing the developer effort and the time cost in manual inspection of source code when attempting to identify relevant buggy code files or functions. Automation is particularly essential for large software projects which are flooded by bug reports that must each be mapped to a relevant source code file among the thousands forming the software project. To address the challenge of automating localization, the research community has recently investigated various information retrieval (IR) techniques~\cite{salton1986introduction,frakes1992information,blei2003latent,schutze2008introduction,deerwester1990indexing,manning1999foundations}.
In the proposed tools~\cite{zhou2012should,wen2016locus,youm2015bug,wong2014boosting,saha2013improving,wang2014version,lukins2010bug}, information tokens are extracted from a given bug report to formulate a query to be matched in a search space of documents formed by the collections of source code files and indexed through tokens extracted from source code properties. IR-based bug localization (IRBL) tools then rank the documents based on a probability of relevance (often measured as a similarity score). Highly ranked files are predicted to be the ones that are likely to contain the buggy code. This process is thus expected to reduce the number of files on which a developer must focus her examination. 

Despite growing interest in the literature, with numerous approaches continuously claiming new performance improvements over the state-of-the-art, we are not aware of any adoption in the developer community, nor any integration in other research approaches such as automated repair. As demonstrated by a recent study by Lee et al.~\cite{lee2018bench4bl}, this is mainly due to:

\begin{enumerate}[leftmargin=*]
	\item {\em A limited performance from the state-of-the-art}: to date, empirical assessments in the literature~\cite{lee2018bench4bl} indicate Mean Average Precision metrics between 0.35 and 0.38 and Mean Reciprocal Rank metrics between 0.43 and 0.52. Concretely, even the best performing IRBL tools still fail every other time to adequately associate the bug reports with the relevant source code files.
	\item {\em A lack of comprehensive validation of the value of the various  IR features\footnote{Generally, features are derived from tokens. As a result, for convenience the terms features and tokens are used interchangeably in this paper.}}: literature approaches incrementally add new dimensions of comparison by considering additional information features or by changing the weight of different information tokens. Unfortunately, to date, the community still lacks a clear overview of the information gain that each feature retrieved through IR contributes to the localization process.
\end{enumerate}

{\bf This paper.} Towards contributing to pushing the frontiers of IR-based bug localization (IRBL) further, we propose to undertake a comprehensive investigation on the information gain provided by a variety of features that are commonly extracted from bug reports and source files for IRBL tasks. 
By further correlating the use of specific feature sets with the performance of different state-of-the-art tools, we are able to establish the need for building an IRBL approach where the weight of similarity scores between bug report features and source code features are learned for different specific groupings of bug reports. 
We refer to it as a ``divide-and-conquer'' (\toolname) strategy, which offers an opportunity to substantially improve the overall performance of IRBL in large-scale experiments.

In a typical IRBL tool, given a bug report and the source code files of a project, the weights associated to the similarity scores between extracted features are statically set and will be the same for all bug reports/source code files. In contrast, \toolname aims to dynamically select the weights that must be used for a given pair (bug report/source code file) following a training phase which learns what kinds of information tokens are important to these ``kinds'' of bug reports/files pairs. We approximate the ``kind'' of bug reports/file pairs by building on training sets that are successfully and exclusively localized by existing state-of-the-art tools.
 In summary, this paper presents the following main contributions to the research community on IRBL:
\begin{itemize}
	\item We dissect the query formulation ({\em i.e., what information tokens from bug reports are used to search for relevant buggy files by matching appropriate information tokens from source code}) in state-of-the-art IRBL systems. Then we assess the contribution of different information tokens on the localization performance. Our experiments compare six state-of-the-art IRBL tools leveraging an extensive database collected for Bench4BL~\cite{lee2018bench4bl}. We observe that different groups of bug reports that have been successfully localized by all or at least one or only one state-of-the-art tool have distinctive attributes which make specific information tokens more or less significant, in terms of contribution to the localization performance. 
	\item We propose a learning approach to improve the performance of IRBL. The key idea of the approach (named \toolname) is based on the finding that each state-of-the-art tool appears to be providing good localization performance on a specific set of bug reports where others fail. The gradient boosting learning algorithm is leveraged to capture the weight that the different information tokens have when the association of a bug report with a source code file, from a specific subset of the dataset, is a successful localization. We then build a multi-classifier where prediction probabilities by different classifiers are combined and reordered to yield the IRBL ranked list of potentially buggy files.
	\item We extensively assess \toolname using Bench4BL~\cite{lee2018bench4bl}, the largest and most comprehensive benchmark that was recently proposed in the literature. The data are from 45 projects (amounting to 5\;321 bug reports and 70\;675 java files). We empirically show that the proposed \toolname approach outperforms the state-of-the-art to yield record Mean Average Precision and Mean Reciprocal Rank values for IRBL, respectively at 0.52 and 0.63. \toolname is further able to localize 50\% of bugs at Top1, 77\% at Top5 and 85\% at Top10.
\end{itemize}

 The remainder of this paper is organized as follows: we first  detail background information on tools used and features leveraged in IR-based localization in Section~\ref{sec:background}. The empirical study for dissecting IRBL performance is described in Section~\ref{sec:study}. Our \toolname approach is presented in Section~\ref{sec:approach} and evaluated in Section~\ref{sec:evaluation}.  We provide final remarks about our work in Section~\ref{sec:discussion}, discuss related work in Section~\ref{sec:related} and conclude in Section~\ref{sec:conclusion}.

\section{Information Features Used in IRBL Tools}
\label{sec:background}

In the context of IR-based bug localization (IRBL), each bug report is treated as a query while the source files in a project form a document collection (i.e., the target search space).
Since the performance of IR systems is generally limited by the linguistic variations present in natural language texts~\cite{schutze2008introduction}, a classic IR challenge lies in effectively recognizing the features in the query and document~\cite{greengrass2000information,frakes1992information}.
Towards providing state-of-the-art tools for IRBL, researchers have investigated a variety of information tokens that can be identified in bug reports and source code files.
We conducted a quick literature review in order to identify which features (i.e., information tokens) are considered by state-of-the-art tools. We consider the state-of-the-art tools that have been studied by Lee et al.~\cite{lee2018bench4bl} in a recent comprehensive reproduction study.

\begin{figure}[!h]
\resizebox{\linewidth}{!}{%
	\includegraphics[width=0.75\textwidth]{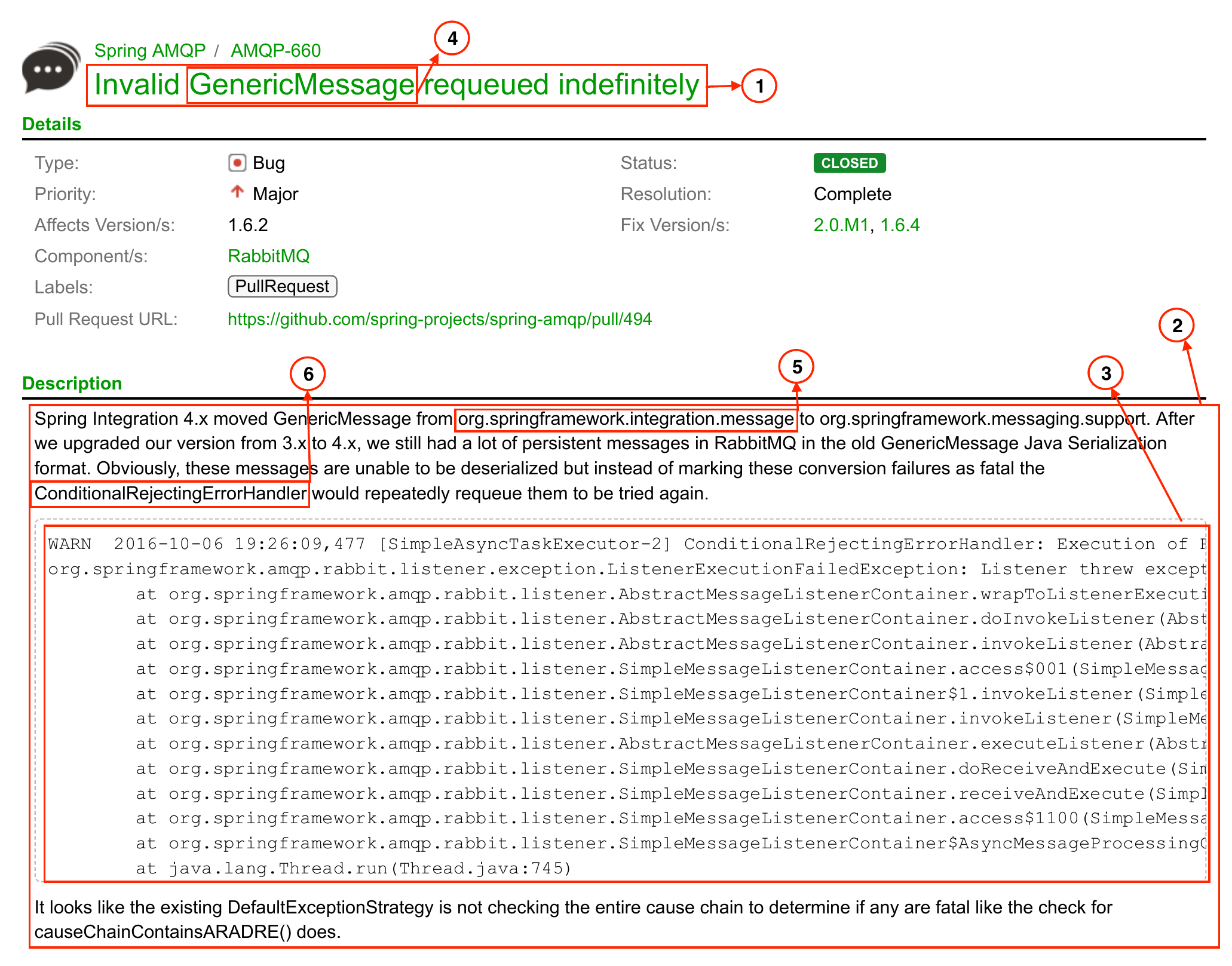}
	}
\caption{Example bug report with (1) Summary, (2) Description, (3) Stack Trace, (4) Summary Hint, (5) Description Hint, and (6) Code Element.}
\label{fig:sampleBugReport}
\end{figure}

A bug report, such as the one illustrated in Figure~\ref{fig:sampleBugReport}, is generally submitted after encountering an issue while running a software program, and typically provides a description of a failure. It is then stored in a bug tracking system for investigation by project developers.
The bug tracking system then records the time at which the bug was reported, the identity of the person who reported it, as well as other information related to the severity or the affected software version. Occasionally, the bug description may include information on the erroneous program behaviour, and the details on how to reproduce the bug hint at the location of the fault in the code (in form of code blocks or stack traces). These information provided with the bug reports can be processed to extract relevant features that could be relevant for implementing IRBL.
For example, code-related terms such as package names and class names found in the summary and description, in addition to stack traces and code blocks, as separate features referred to as {\em hints}.

In general, state-of-the-art tools to IRBL develop specific strategies towards ensuring that queries are processed adequately to find the information that allows accurate matching with source code file information.
Nevertheless, a key impactful factor in the performance of the approach remains the features that are extracted as representative and discriminating information of bug reports and source code files. In order to identify which features are considered in the literature, we provide the following summary information for recent tools:

\begin{itemize}[leftmargin=*]
	\item Zhou et al.~\cite{zhou2012should} have initiated the breakthrough in IRBL by radically raising the precision to about 50\%. The tool merely treats source code as text to match with natural language text of bug reports. Moreover, it leverages the similarity among bug reports to guide localization, and uses file sizes to weight probability scores (given that larger files are more likely to include bugs).
	\item Saha et al.~\cite{saha2013improving} proposed to treat separately summary and description parts of a bug report. They further extracted specific information from source code files into a structured format (class names, variable names, comments) to improve matching.
	\item Wang et al.~\cite{wang2014version} combined the works of Zhou et al.~\cite{zhou2012should} and Saha et al.~\cite{saha2013improving} and further considered version history to improve prediction (a previously buggy file is likely to still contain bugs). They later extended their work to consider reported information~\cite{wang2016Amalgam+}.
	\item Wong et al.~\cite{wong2014boosting} proposed to segment source code files into smaller segments and used stack trace information to improve bug localization.
	\item Youm et al.~\cite{youm2015bug}, in addition to information tokens such as stack traces, method names, and similarity among bug reports and method names, further consider method level matching and exploit comments from bug reports.
	\item Wen et al.~\cite{wen2016locus} have focused on the change level in source code and attempted to separate natural language tokens from code entities to improve matching between the bug report and source code.
\end{itemize}

In this work, we consider extracting such common features which can be reliably and readily computed in a large scale dataset (e.g., we do not consider similarity among bug reports because of the combinatorial explosion of pairwise combinations), and which are available once a bug report is submitted (e.g., we do not consider comments which may be subsequently added to the bug reports). Overall, we focus in this study on :
\begin{enumerate}[leftmargin=*]
	\item {\em Textual information of bug reports and source code files}: source code files that are textually similar to the bug report text tend to be associated to the reported bug~\cite{lukins2010bug}.
	\item {\em Structured information from source code}:  different fields in code (e.g., class names, package names, comment, etc.) have varying importance for matching bug report vocabulary~\cite{moreno2013relationship,saha2013improving}.
	\item {\em Structured information from bug reports}: different parts of bug reports, such as title and body, may contain specific or verbose information for matching. Furthermore, some code elements can often be identified in bug reports, which could be more effective for bug localization~\cite{saha2013improving,wong2014boosting,youm2015bug}.
	\item {\em Stack traces}: the bug location is likely among the classes or methods listed in the stack trace~\cite{schroter2010stack, wong2014boosting}.
	\item {\em Segmentation}: matching at the code hunk level~\cite{wen2016locus} or dividing source code files into equally sized segments ~\cite{wong2014boosting} can provide more accuracy in localizing bugs~\cite{ye2014learning}.
	\item {\em Commit log}: messages included in source code version management systems can provide the description of functionalities that match user bug report text better than source code tokens~\cite{wen2016locus}.
\end{enumerate}

The detailed list of features considered in this paper is presented in Section~\ref{sec:dissection}.

\section{Empirical Study on IRBL tools}
\label{sec:study}

In this section, we describe the setup and results of a large empirical study that we have conducted to investigate the impact of different IRBL features as well as the differences in performances by current state-of-the-art tools. Our objectives are to assess the impact of the query formulation on the performance, and to provide comprehensive insights into the value of different IR features for bug localization.
We recall that this study is focused on working tools available to the community. We refer the reader to the study of Thomas et al.~\cite{thomas2013impact} on the impact of classifier configuration, which focuses on the underlying IR methods. 

\subsection{Research Questions}
Our study focuses on the following questions:
\begin{enumerate}
	\item[] {\bf RQ-1}: {\em Are state-of-the-art tools diversely successful depending on the samples of the benchmark?}
	A recent reproducibility study has shown that current tools have an overall similar performance in terms of average precision~\cite{lee2018bench4bl}. However, the authors did not assess with the large dataset of Bench4BL~\cite{lee2018bench4bl} whether these
	  tools have affinities for specific sets of bug reports/code files.
	\item[] {\bf RQ-2}: {\em Which combinations of features provide the best information gain for IR-based bug localization?}
	 Although the literature recurrently adds new features that are expected to improve overall localization, little
	  knowledge has been established by the community on the actual contribution of each feature, and whether this contribution varies depending on the project.
\end{enumerate}

\subsection{Experiment Setup}
For the purpose of our study, we consider six state-of-the-art tools which are broadly used in the literature. Although some works have been later extended by tweaking some parameters and features, we consider the originally-published work and the associated implementation details to perform our study. Table~\ref{tab:techs} enumerates the tools of which the implementations were readily available and have been applied to the Bench4BL benchmark by Lee et al~\cite{lee2018bench4bl}.

\begin{table}[!h]
\centering
		\scriptsize
\caption{Tools considered in this study.}%
\label{tab:techs}
\resizebox{1\linewidth}{!}{%
		\begin{tabular}{lrrr}
			Name &  Venue  & Year \\
        \noalign{\smallskip}\hline\noalign{\smallskip}
        BugLocator~\cite{zhou2012should} & Intl. Conf. on Software Engineering & 2012\\
       Bluir~\cite{saha2013improving} & Intl. Conf. on Automated Software Engineering & 2013\\
       Amalgam~\cite{wang2014version} & Intl. Conf. on Program Comprehension & 2014\\
       Brtracer~\cite{wong2014boosting} & Intl. Conf. on Software Maintenance and Evolution & 2014\\
       Blia~\cite{youm2015bug} & Asia-Pacicific Software Engineering Conference& 2015\\
       Locus~\cite{wen2016locus} & Intl. Conf. on Automated Software Engineering & 2016\\
			\hline\noalign{\smallskip}
    \end{tabular}
        }
 \end{table}

\subsection{Dataset}
To conduct our study, we exploit the dataset and benchmark provided by Bench4BL~\cite{lee2018bench4bl}.
This benchmark was recently proposed by Lee et al. in an effort to push the assessment of current tools.
Bench4BL (i.e., a benchmark for Bug Localization) was then leveraged to perform a comprehensive reproduction study
on state-of-the-art IRBL tools. Table~\ref{tab:dataset} enumerates the projects available in the Bench4BL benchmark.
For the purpose of our study, we have thoroughly investigated the datasets and applied further constraints to obtain a clean dataset.

For each dataset, we have performed an extra curation step by ensuring that all files tagged as fixing a given bug are still available in the latest code version of the Git repository (as of December 2018).
When any one file of them is not available, we discard the associated bug report from our experiments. Eventually, our experiments are done on 5\,321 bug reports filed in 45 projects (whereas BenchBL originally includes 8\,652 reports from 46 projects).

In order to identify bug-fixing patches, Bench4BL leverages the bug linking strategies enforced when developers use the JIRA bug tracking system.
Bug tracking systems are crawled and bug links are verified based on two checks: i) Bench4BL checked for explicit commit ids (i.e., Git hashes) and file paths associated to the bug on the bug tracking database: for each file impacted by an identified commit, it considers the corresponding change as a bug fix change. ii) Similarly, Bench4BL also checked commit logs to identify bug report ID and associate the corresponding changes as bug fix changes. Finally, the Bench4BL dataset is curated by selecting only bug reports that are indeed considered as such and are thus resolved and tagged as RESOLVED or FIXED, and completed with status CLOSED. Eventually, the {\em cleaned Bug reports} amount to 5321 (second column of Table~\ref{tab:dataset}).

For the assessment of our proposed approach\footnote{The empirical study part is done with all bug reports as we are blindly investigating the importance of bug report features (whether future or past data)}, we further clean the data from all bug reports that are suspected of being post-fix activities.
We consider such bug reports to represent future data and may thus lead to artificial performance. For example, some bug reports are submitted by developers to keep track of what the code changes are meant to correct. The associated descriptions are too precise and may unrealistically match the source code (e.g., with file names and method names) with very high accuracy. Concretely we dismiss cases where the bug reporter and bug fixer (change committer) are same. This equality is controlled via the email. We also remove bug reports cases where a patch attachment is provided by the reporter or via a comment within the hour. Eventually, the {\em pre-fix bug reports} amount to 4005 (fifth column of Table~\ref{tab:dataset}) Table~\ref{tab:dataset} reports all the statistics of the datasets. In the end, our experiments are done with similarity computations for over 35 millions bug report-source code file pairs.



\begin{table}[!t]
\centering
		\scriptsize

\caption{Descriptive Statistics of Curated Bench4BL.}%
\label{tab:dataset}
\resizebox{\linewidth}{!}{%
		\begin{tabular}{lrrrrrr}
			\hline\noalign{\smallskip}
			Project & \makecell[c]{\# Cleaned\\ Bug Reports} & \makecell[c]{\# with \\ Same Email} & \makecell[c]{\# with \\ Attachment} & \makecell[c]{\# Pre-fix \\ Bug Reports} &\makecell[c]{\# Source \\Code Files} & \makecell[c]{\# Bug \\Report-Source}\\
			
        \noalign{\smallskip}\hline\noalign{\smallskip}

APACHE-CAMEL			&	1114	&	128	&	182	&	797	&	18671	&	20799494	\\
APACHE-CODEC			&	39		&	5	&	7	&	28	&	126	&	4914	\\
APACHE-COLLECTIONS		&	32		&	0	&	5	&	26	&	535	&	17120	\\
APACHE-COMPRESS			&	99		&	16	&	28	&	56	&	354	&	35046	\\
APACHE-CONFIGURATION	&	13		&	0	&	1	&	11	&	458	&	5954	\\
APACHE-CRYPTO			&	1		&	0	&	0	&	1	&	87	&	87	\\
APACHE-CSV				&	12		&	1	&	3	&	8	&	31	&	372	\\
APACHE-HBASE			&	452		&	73	&	168	&	228	&	3758	&	1698616	\\
APACHE-HIVE				&	727		&	156	&	234	&	395	&	6200	&	4507400	\\
APACHE-IO				&	78		&	12	&	10	&	43	&	246	&	19188	\\
APACHE-LANG				&	144		&	31	&	14	&	99	&	324	&	46656	\\
APACHE-MATH				&	17		&	4	&	1	&	4	&	1324	&	22508	\\
APACHE-WEAVER			&	2		&	1	&	0	&	1	&	89	&	178	\\
JBOSS-ELY				&	19		&	0	&	0	&	5	&	936	&	17784	\\
JBOSS-ENTESB			&	6		&	0	&	0	&	6	&	23	&	138	\\
JBOSS-JBMETA			&	11		&	0	&	1	&	10	&	852	&	9372	\\
JBOSS-SWARM				&	38		&	0	&	2	&	35	&	1358	&	51604	\\
JBOSS-WFARQ				&	1		&	0	&	0	&	1	&	171	&	171	\\
JBOSS-WFCORE			&	321		&	0	&	11	&	310	&	4141	&	1329261	\\
JBOSS-WFLY				&	519		&	0	&	55	&	461	&	8581	&	4453539	\\
JBOSS-WFMP				&	3		&	0	&	0	&	3	&	84	&	252	\\
SPRING-AMQP				&	77		&	7	&	3	&	67	&	453	&	34881	\\
SPRING-ANDROID			&	7		&	0	&	0	&	7	&	305	&	2135	\\
SPRING-BATCH			&	253		&	1	&	27	&	239	&	1897	&	479941	\\
SPRING-BATCHADM			&	15		&	0	&	0	&	15	&	231	&	3465	\\
SPRING-DATACMNS			&	119		&	0	&	8	&	110	&	745	&	88655	\\
SPRING-DATAGRAPH		&	3		&	1	&	0	&	2	&	287	&	861	\\
SPRING-DATAJPA			&	119		&	0	&	16	&	103	&	367	&	43673	\\
SPRING-DATAMONGO		&	213		&	0	&	17	&	195	&	821	&	174873	\\
SPRING-DATAREDIS		&	41		&	3	&	2	&	37	&	720	&	29520	\\
SPRING-DATAREST			&	77		&	1	&	2	&	72	&	419	&	32263	\\
SPRING-LDAP				&	42		&	2	&	6	&	29	&	529	&	22218	\\
SPRING-MOBILE			&	11		&	0	&	0	&	11	&	105	&	1155	\\
SPRING-ROO				&	130		&	0	&	17	&	113	&	1077	&	140010	\\
SPRING-SEC				&	251		&	0	&	20	&	242	&	2304	&	578304	\\
SPRING-SECOAUTH			&	18		&	3	&	1	&	14	&	760	&	13680	\\
SPRING-SGF				&	30		&	8	&	5	&	19	&	902	&	27060	\\
SPRING-SHDP				&	40		&	26	&	0	&	14	&	1068	&	42720	\\
SPRING-SOCIAL			&	13		&	0	&	0	&	35	&	217	&	2821	\\
SPRING-SOCIALFB			&	12		&	1	&	0	&	11	&	261	&	3132	\\
SPRING-SOCIALLI			&	4		&	0	&	0	&	4	&	183	&	732	\\
SPRING-SOCIALTW			&	8		&	0	&	1	&	7	&	156	&	1248	\\
SPRING-SPR				&	31		&	0	&	5	&	25	&	6906	&	214086	\\
SPRING-SWF				&	53		&	0	&	7	&	40	&	748	&	39644	\\
SPRING-SWS				&	106		&	1	&	10	&	66	&	865	&	91690	\\
\noalign{\smallskip}\hline
Total	&	5321	&	481	&	869	&	4005 	&	70675	&	35088421	\\

		\end{tabular}
		}
\end{table}

\subsection{Performance Metrics}
Assessment is quantified with common metrics used in the literature, with a focus on Mean Average Precision (MAP) and Mean Reciprocal Rank (MRR).

\noindent
{\em Precision} is a measure of accuracy in bug localization that shows how many files are correctly recommended within a given TopN files.
\begin{equation}
\mbox{P}(n) = \frac{\# of\ buggy\ files\ at\ top\ n} {n}
\end{equation}

\noindent
{\em Recall} is a measure of coverage in bug localization that shows how many files are correctly recommended within given TopN files over the actually fixed files by a developer for a given bug report.
\begin{equation}
\mbox{R}(n) = \frac{\# of\ buggy\ files\ at\ top\ n} {\# of\ actual\ fixed\ files}
\end{equation}

\noindent
{\em Average Precision} is computed for a given bug report by aggregating precision values of several positively recommended files.
\begin{equation}
AP = \sum_{i=1}^{N} = \frac{\mbox{P}(i) \times \mbox{pos}(i)} {\# of\ positive\ instances}
\end{equation}
where $N$ is the number of ranked files by an IRBL tool, i is a rank in the ranked list of recommended files. $\mbox{pos}(i)$ indicates whether the $i_{th}$ file in the ranked list is a buggy file (i.e., $\mbox{pos}(i) \in 0, 1$).

\begin{itemize}

\item \textbf{Mean Average Precision (MAP)} is computed by taking the mean value of AP values across all bug reports:

\begin{equation}
MAP = \frac{1}{M} \sum_{j=1}^{M} \mbox{AP}(j)
\end{equation}

where $M$ is the number of all bug reports and $\mbox{AP}(j)$ is the
average precision of bug report $j$.

\item \textbf{Mean Reciprocal Rank (MRR)}  computes
the mean value of the position of the first buggy file in the ranked list recommended an IRBL tool as follows:

\begin{equation}
MRR = \frac{1}{M} \sum_{j=1}^{M} \frac{1}{rank_j}
\end{equation}

where $M$ is the number of all bug reports and $rank_i$ means the position of the first buggy file in the ranked list for $i_{th}$ bug report.

\item \textbf{Top-N Rank (TopN)} computes the number of bug reports having at least one relevant file in the first N files of the retrieved list. 
\begin{equation}
\label{eq:success}
    TopN(r) = \begin{cases}
      & \mbox{report $r$ having at least} \\
    1 & \mbox{one relevant file in the within}\\
      & \mbox{top-$N$ files recommended} \\
    0 & \mbox{otherwise}
\end{cases}
\end{equation}

\begin{equation}
\label{eq:topN}
    TopN = \sum_{r=1}^{R} TopN(r)
\end{equation}

where $r\in R$ is a bug report, and $N\in \mathbb{N}$
is a parameter of
how many recommendations (i.e., files to fix) to look up from the results of $f$.
Let  $R$ be a set of bug reports, and $\mathbb{N}$ 1, 5, 10 respectively. We also present a version of this metric in terms of percentage against the total number of bug reports that must be validated.
\end{itemize}

\begin{figure*}[!t]
\minipage[t]{0.33\textwidth}
		\caption*{Top1.}
		\includegraphics[width=1\linewidth]{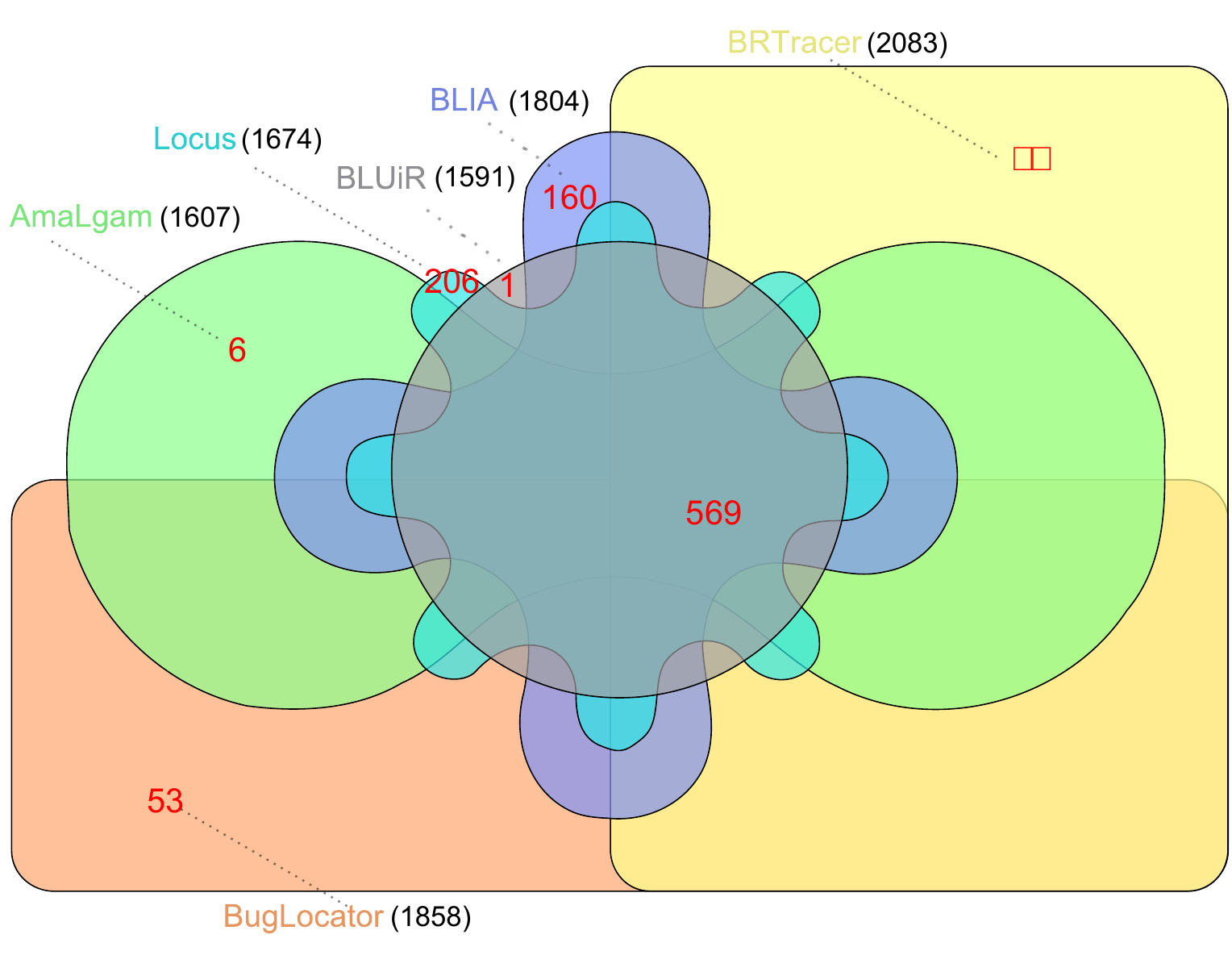}
\endminipage\hfill
\minipage[t]{0.33\textwidth}
		\caption*{Top5.}
		\includegraphics[width=1\linewidth]{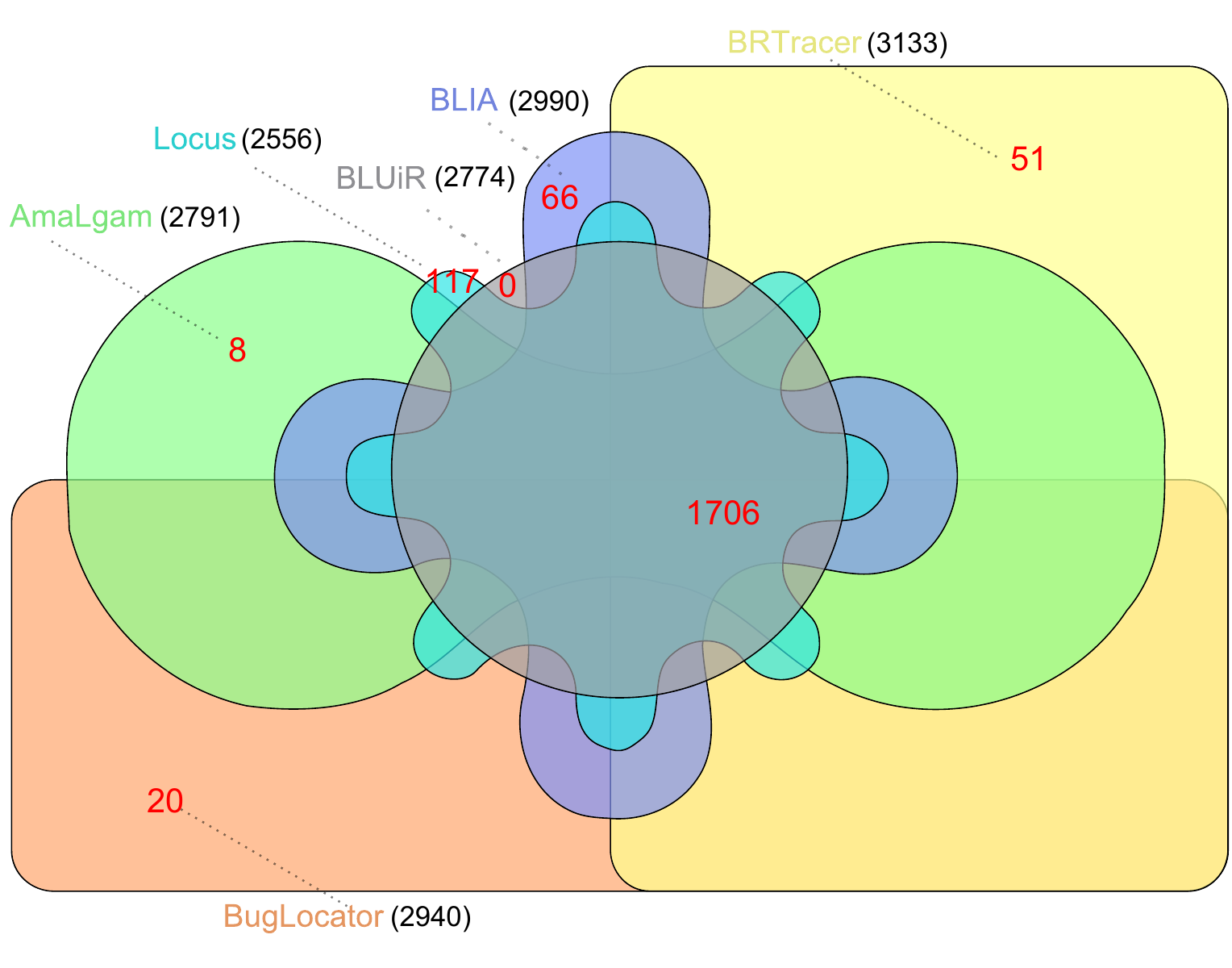}
\endminipage\hfill
\minipage[t]{0.33\textwidth}
		\caption*{Top10.}
		\includegraphics[width=1\linewidth]{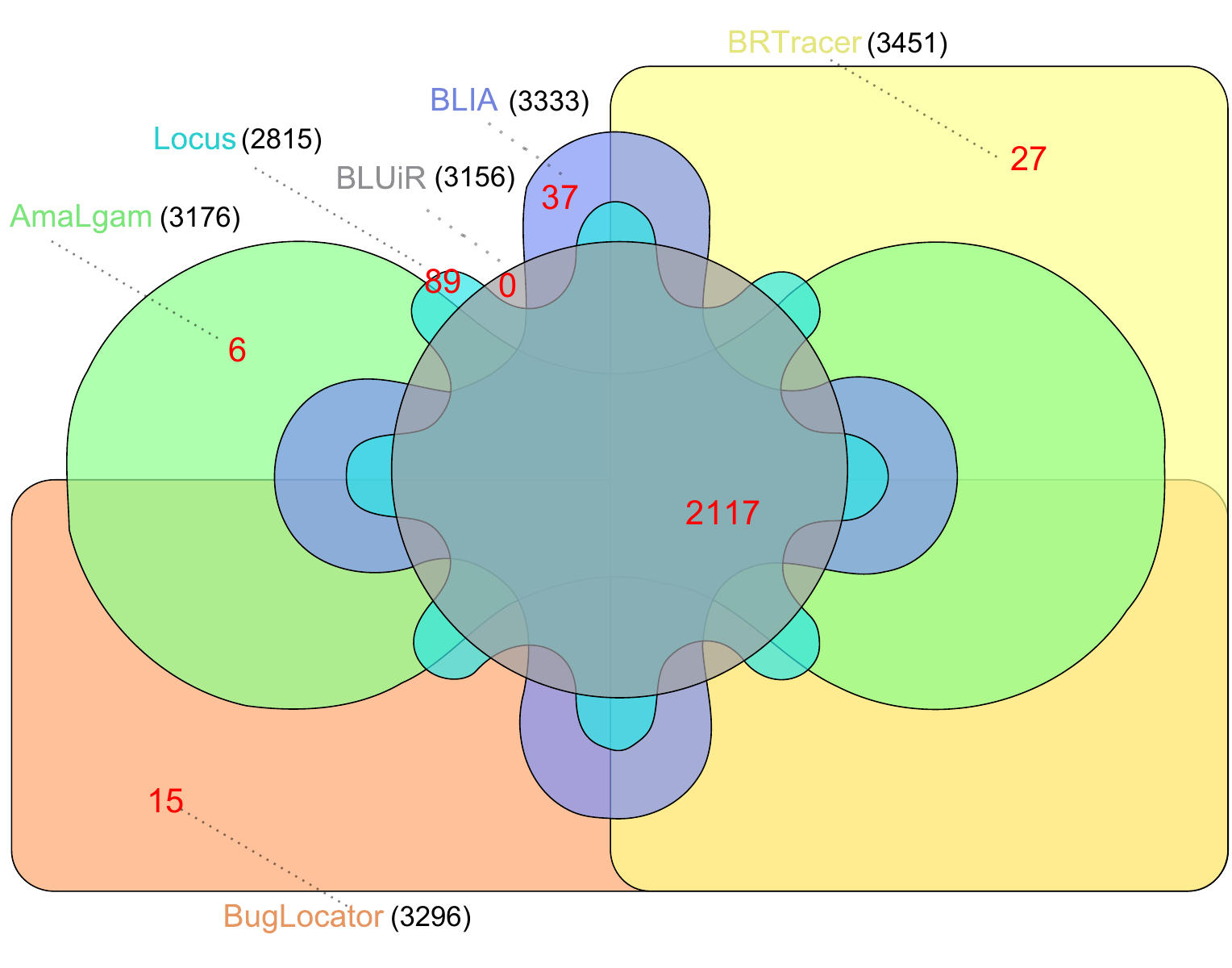}
\endminipage\hfill
\caption{Successful recommendations of IRBL tools and the overlapping among them for Top1, Top5 and Top10.}
\label{fig:venns}
\end{figure*}

\subsection{RQ-1: Affinities among state-of-the-art tools} \label{rq1}

To answer our first research question (RQ1), we perform the overlap analysis 
of IRBL tools and investigate whether they show some complementary
relationships with regards to being successful for specific sets of bug reports. Potential outcomes of this study are (1) all the tools are successful only for a specific set of bug reports so that we cannot identify any factor affecting the performance of each tool or
(2) different tools are successful for each different set of bug reports so that we can identify
what properties affect the performance of each individual tool.

\subsubsection{Design}
\label{sec:rq1design}

In this study, we consider the execution results provided in the reproduction study of Lee et al.\cite{lee2018bench4bl} on Bench4BL.
For each tool listed in Section~\ref{tab:techs}, we collect its results
(i.e., ordered lists of files suspected as bug locations)
for every bug report of the projects listed in our dataset (cf. Table~\ref{tab:dataset}).

We then identify which tool successfully recommends files to fix for each bug report shown in Table~\ref{tab:dataset}.
Since the results of IRBL tools are ordered lists of files, we take Top1, Top5 and Top10
recommended files when computing the performance of each tool.
In this study, we classify whether an IRBL tool is successful for
a specific bug report using the Equation~\ref{eq:topN}. Figure~\ref{fig:venns} presented with Venn diagrams~\cite{heberle2015interactivenn} illustrates the relationships among state-of-the-art tools with respect to the number of bug reports that they can be localized exclusively or not at different positions (Top1, Top5, and Top10). For example, when we consider the diagram for Top1, BLIA makes correct localization for 1804 bug reports, among which \textcolor{red}{160} are localized correctly only by BLIA at Top1 position.

The intersection among all tools tends to be larger when we increase $N$ (of Top$N$).
This indicates that recommendations by different IRBL tools gradually
converge to a consensus as $N$ is larger. However, increasing $N$ inevitably
results in more false positives and imprecise recommendations.

Although several bugs are correctly localized simultaneously by several tools, there is still a notable portion of the bugs that are localized exclusively by each of the tools.
13.0\% ( 524 = 6+206+1+160+98+53), 6.6\% ( 262 = 8+117+0+66+51+20), and 4.6\% ( 174 = 6+89+0+37+27+15) of bug reports are successfully localized by
only a single tool at Top1, Top5, and Top10 respectively. This implies that
some tools appear to be exclusively successful for a specific subset of bug reports.
If we can figure out the properties of the bug reports/source code files that make a specific tool perform better, we can tune the IRBL to fit the localization decision depending on the bug report/source code file pair that is being assessed.

%


We perform an overlap analysis~\cite{oliveto2010equivalence} between the tools for the bug reports that are localized correctly.
The analysis is based on the following Equation~\ref{eq:overlap}:  
\begin{equation}
\label{eq:overlap}
\begin{aligned}
    C_{A \cap B} = \frac{\vert C_{A} \cap C_{B} \vert}{\vert C_{A} \cup C_{B} \vert} \% \\
    C_{A \setminus B} = \frac{\vert C_{A} \setminus C_{B} \vert}{\vert C_{A} \cup C_{B} \vert} \% \\
    C_{B \setminus A} = \frac{\vert C_{B} \setminus C_{A} \vert}{\vert C_{B} \cup C_{A} \vert} \%
\end{aligned}
\end{equation}
where $C_{A}$ represents the bug reports that have been correctly localized by tool $A$. $C_{A \cap B}$ is the percentage of the overlap between the sets of bug reports that have been correctly localized, while $C_{A \setminus B}$ is the percentage of the bug reports whose localization is correctly found by $A$ but missed by $B$. 




Table~\ref{tab:overlap} shows the overlap analysis. The results quantitatively confirms the visual conclusions from the Venn diagrams of Figure~\ref{fig:venns}.

\begin{table}[!!h]
\centering
	\setlength\tabcolsep{0.5pt}
	\caption{Overlap analysis results.}
	\label{tab:overlap}
\resizebox{1.0\linewidth}{!}{
\begin{tabular}{ll|rrr|rrr|rrr}

\toprule
\multicolumn{2}{c}{\bf Tool} & \multicolumn{3}{c}{\bf Top1} & \multicolumn{3}{c}{\bf Top5} & \multicolumn{3}{c}{\bf Top10} \\\cline{0-10}

    A & B & $C_{A \cap B}$ & $C_{A \setminus B}$ & $C_{B \setminus A}$ & $C_{A \cap B}$ & $C_{A \setminus B}$ & $C_{B \setminus A}$ & $C_{A \cap B}$ & $C_{A \setminus B}$ & $C_{B \setminus A}$  \\
\midrule
 BugLocator &     Bluir &  49.8 &  30.9 &  19.3 &  73.5 &  15.8 &  10.7 &  80.0 &  11.9 &   8.0 \\
 BugLocator &   Amalgam &  49.9 &  30.5 &  19.6 &  73.5 &  15.5 &  11.0 &  80.1 &  11.6 &   8.3 \\
 BugLocator &  Brtracer &  74.4 &   7.8 &  17.8 &  86.9 &   3.6 &   9.5 &  90.5 &   2.6 &   6.9 \\
 BugLocator &      Blia &  53.7 &  24.3 &  22.0 &  78.1 &  10.2 &  11.7 &  84.0 &   7.5 &   8.5 \\
 BugLocator &     Locus &  46.6 &  30.5 &  22.9 &  64.2 &  23.7 &  12.2 &  68.6 &  22.3 &   9.1 \\
      Bluir &   Amalgam &  98.8 &   0.1 &   1.1 &  99.3 &   0.0 &   0.6 &  99.4 &   0.0 &   0.6 \\
      Bluir &  Brtracer &  48.5 &  15.8 &  35.7 &  72.8 &   8.3 &  18.8 &  79.8 &   6.1 &  14.1 \\
      Bluir &      Blia &  44.7 &  23.1 &  32.2 &  72.7 &  10.4 &  16.9 &  79.9 &   7.6 &  12.5 \\
      Bluir &     Locus &  40.1 &  28.2 &  31.7 &  60.2 &  23.2 &  16.6 &  66.2 &  21.7 &  12.2 \\
    Amalgam &  Brtracer &  48.7 &  16.1 &  35.3 &  72.8 &   8.6 &  18.6 &  79.8 &   6.4 &  13.8 \\
    Amalgam &      Blia &  44.8 &  23.4 &  31.8 &  72.7 &  10.7 &  16.6 &  80.0 &   7.9 &  12.2 \\
    Amalgam &     Locus &  40.2 &  28.5 &  31.3 &  60.1 &  23.5 &  16.4 &  66.2 &  21.9 &  11.9 \\
   Brtracer &      Blia &  57.4 &  26.9 &  15.6 &  80.7 &  11.7 &   7.5 &  86.8 &   8.2 &   5.0 \\
   Brtracer &     Locus &  49.3 &  33.5 &  17.2 &  65.1 &  25.8 &   9.1 &  69.4 &  23.9 &   6.7 \\
       Blia &     Locus &  42.9 &  31.2 &  25.9 &  63.1 &  24.8 &  12.1 &  67.8 &  23.2 &   9.0 \\
\end{tabular}
}
\end{table}

\find{
\label{find:tendToAgree}
 State-of-the-art tools tend to converge on suggesting the same locations for many bug reports when the list of recommendations is extended. Detailed comparisons among tools also reveal that some bug reports sets appear to be more localizable by specific tools.
}

\subsection{Feature Engineering}
\label{sec:dissection}


Feature engineering is an important process for IRBL: tools in the literature mainly differentiate from each other on the variety of features that are considered. Indeed, the overall assumption is that the IRBL query formulation ({\em i.e., what information tokens from bug reports are used to search for relevant buggy files by matching appropriate information tokens from source code}) is one of the essential steps in bug localization. 
Nevertheless, there is scarce knowledge about what are the most impactful query formulation schemes. To answer the second research question (RQ2), we investigate in this paper which information tokens are useful for an efficient query formulation. The objective is thus to assess the contribution of different combinations of tokens, across bug reports and source code files, in the success of bug localization.

\begin{figure}[!h]
\resizebox{\linewidth}{!}{%

  \includegraphics[width=\textwidth]{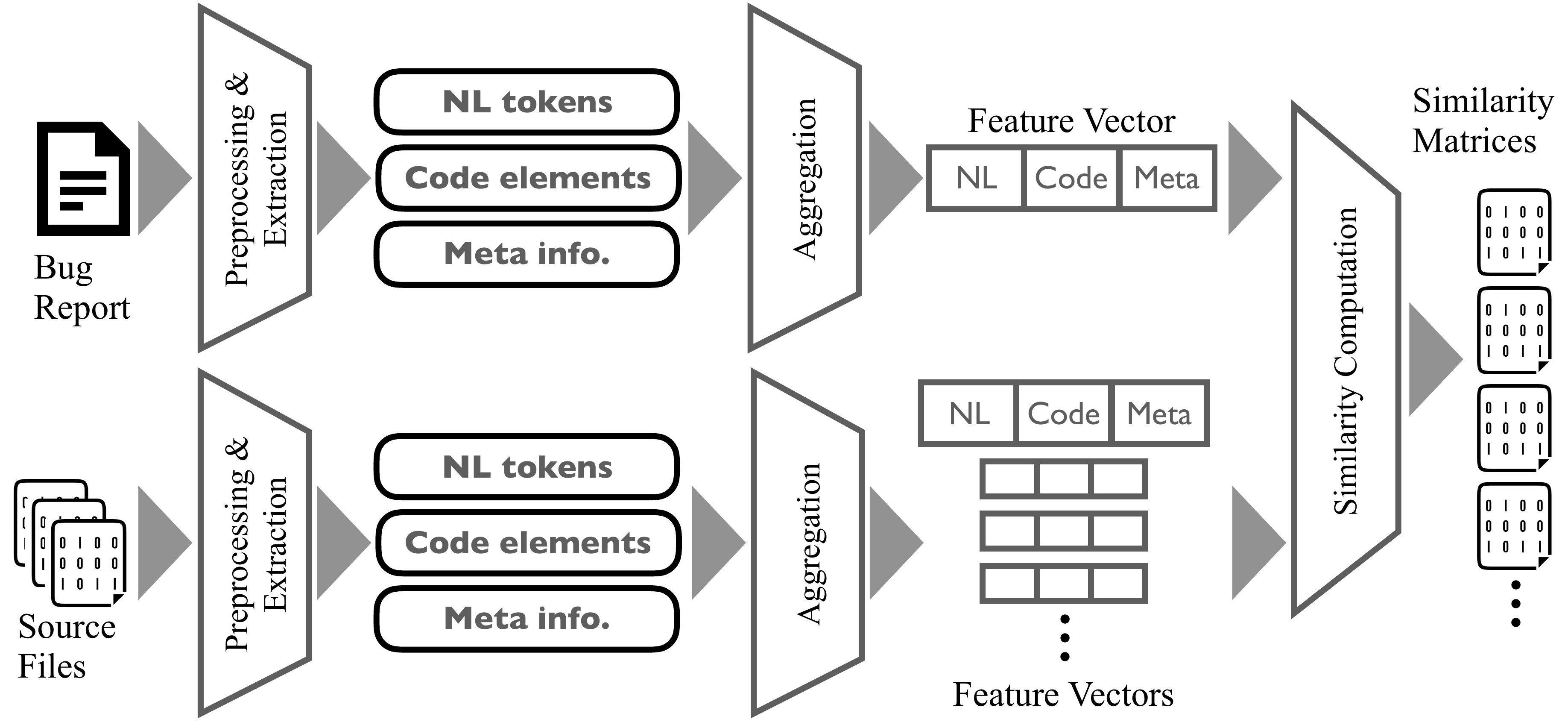}
	}
\caption{Feature engineering process.}
\label{fig:archi}
\end{figure}

To avoid biases of heuristics that various state-of-the-art tools have developed to boost performance in localization, we implement a generic approach based on tasks and strategies that are commonly shared by different tools. This generic approach is illustrated in Figure~\ref{fig:archi}. Bug reports and source code files of a project are processed to extract different types of tokens depending on the approach. These include natural language tokens (e.g., bug report description text and source code comments), code elements (e.g., method names in stack traces and source code AST attributes), and other metadata information (e.g., bug report submitter and source code file size). This information tokens are then used to build feature vectors for computing the similarity whose scores are leveraged to produce a ranked list of files that will be recommended for localizing a given reported bug.

\subsubsection{Preprocessing and extraction}
As previously explained in Section~\ref{sec:background}, for our experiments, we have surveyed the features used in the literature and proposed to further refine some of them to more accurately study the impact of different information tokens.
Overall, we consider 10 types of features from source code files (cf. Table~\ref{tab:codeFeatures}) and 7 types of features from bug reports (cf. Table~\ref{tab:bugFeatures} along with the justification of use in bug localization).
\begin{table*}[!t]
\centering
		\scriptsize

\caption{IR features collected from bug reports.}

\label{tab:bugFeatures}
\resizebox{\linewidth}{!}{%
		\begin{tabular}{lll }
			\hline
			\bf{Feature} & \bf{Description} & {\bf Use in bug localization}\\\hline

			summary  &  The summary/title part of the bug report & Usually includes essential keywords about the problem.\\
            description      &   The description part of the bug report & Is generally more verbose and provides additional descriptive tokens\\
            rawBugReport      &   The whole bug report & Contains all textual information\\
            stackTraces      &   The stack traces in the bug report& Entities (classes, files, etc.) in stack traces are likely buggy \\
            codeElements    &   Code snippets in the bug reports& Can be indicative of the code part that is involved\\
            summaryHints    &   Localization hints in the summary& Code-related terms in summary (e.g., package name) is relevant to buggy part \\
            descriptionHints    &  Localization hints in the description & Code-related terms found by parsing description text (e.g., based on camelCase regexp) \\
			 \hline

		\end{tabular}}
\end{table*}

After identifying relevant tokens from bug reports and source code files, the tokenizer proceeds with a lexical analysis following the common steps from the literature, and which has influence in the ultimate performance of the retrieval model~\cite{saha2013improving}: first, text is retrieved and tokens are produced, then stopword removal\footnote{Stop words from NTLK framework: \url{https://www.nltk.org/}} is performed to reduce noise along with programming and project keywords.
Finally, stemming (i.e., PorterStemmer~\cite{karaa2013information}) is applied to all tokens to create homogeneity with the term's root (by conflating variants of the same term).

\begin{table}[!h]
\centering
		\scriptsize

\caption{IR features collected from source code files.}%

\label{tab:codeFeatures}
\resizebox{\linewidth}{!}{%
		\begin{tabular}{ll}
			\hline
			\bf{Feature} & \bf{Description} \\\hline

			packageNames  &  The parsed package names of the source code files\\
            className      &   The parsed class names of the source code files \\
            methodNames      &   The parsed method names of the source code files   \\
            methodInvocation      &   The parsed method invocation of the source code files  \\
            formalParameter    &   The parsed formal parameters of the source code files  \\
            memberReference    &   The parsed member references of the source code files  \\
            documentation    &  The parsed class names of the source code files  \\
            rawSource    &   Source file as a text \\
            hunks    &   The hunks from the commits on the file \\
            commitLogs    &   The commit logs of the file \\
			 \hline

		\end{tabular}
		}
\end{table}

Given that terms from documents can be stack trace terms, code elements or natural language terms, we adopt slight specializations in the preprocessing steps for the different types:

\begin{itemize}[leftmargin=*]

\item Natural language:
      Tokenization is based on white spaces as a classical separator. Tokens are then checked against the WordNet~\cite{miller1995wordnet} dictionary to discard all unknown tokens.
  \item Stack traces and code elements: We use regular expressions to detect stack traces and code elements. Due to the specific nature of stack traces and code elements, the tokenization is based on punctuations, camel case splitting (e.g., findNumber splits into find, number) as well as snake case splitting (e.g {\em find\_number} splits into ``{\em find}, {\em number}'').
  \end{itemize}

\subsubsection{Vectorization and Similarity Computation}
IRBL techniques recurrently treat source code as a form of text on which Natural Language Processing (NLP) techniques can be applied to automatically extract features.
Several tools in the literature build upon the revised Vector Space Model (rVSM)~\cite{zhou2012should} which represents bug reports and source code files as collections of tokens: these documents are then associated in a vector space model where tokens are weighted with term frequencies to calculate the similarity between documents.


In IRBL, we consider all project source code files as constituting the collection of documents  \(d\) forming the search space, and a given bug report as the query \(q\). $\mbox{tf-idf}(t,d)$ assigns to term $t$ a weight in document $d$ that is

\begin{equation}
\label{eq:tfidf}
\mbox{tf-idf}(t,d) = \mbox{tf}(t,d) \times \mbox{idf}(t),
\end{equation}

The document space is then defined as \(D = \{ d_1, d_2, \ldots, d_n \}\) where n is the number of documents in the corpus, $t$ represents the terms, and \mbox{tf} (term frequency) is defined as follows:

\begin{equation}
 \mbox{tf}(t,d) = 1 +\log{(\,{f_{td}}\,)} \newline
\end{equation}

where ${f_{td}}$ refers to the number of occurrences of a term $t$ in document $d$.
The \mbox{idf} (inverse document frequency) is defined by:

\begin{equation}
 \mbox{idf}(t) = \log \left( \frac{1+ n} {1 + \left|\{d : t \in d\}\right|} \right) + 1
\end{equation}
where \(\left|\{d : t \in d\}\right|\) is the number of documents where the term t appears, when the term-frequency function satisfies \(\mathrm{tf}(t,d) \neq 0\). The constant 1 is added to numerator and denominator of the idf, as if an
extra document was seen containing every term in the collection
exactly once, prevents divisions by zero.

Eventually, each document (bug report or source code file) is represented as a vector where each element corresponds to a term in the dictionary (of all terms appearing in the document space), together with a weight given by the $\mbox{tf-idf}$ formula (Equation~\ref{eq:tfidf}). Conservatively, the weight 0 is given to all dictionary terms that do not occur in a document. Ultimately, the vector form is used to calculate the relevance score between a document \(d\) (i.e., a source code file) and a query \(q\) (i.e., a bug report).
This score is computed as the cosine similarity between the associated vector representations:

\begin{equation}
\label{eq:simi}
\mbox{Similarity}(q_n,d_m)=\mbox{cos}(q_n,d_m)= \frac{\vec{V}(q_n)\cdot \vec{V}(d_m)}{\vert\vec{V}(q_n)\vert \vert\vec{V}(d_m)\vert},
\end{equation}

with $\vec{V}(q_n)$ being the vector of term weights of bug report $n$ and $\vec{V}(d_m)$ being the vector of term weights of source code file $m$, where the term weights are computed using Equation~\ref{eq:tfidf}.

\begin{figure*}

\minipage[t]{0.3\textwidth}
	\subcaption*{All bug reports for which \underline{\em at least one} state-of-the-art tool can recommend the correct bug location at Top1 of its list.}
		\includegraphics[width=1\linewidth]{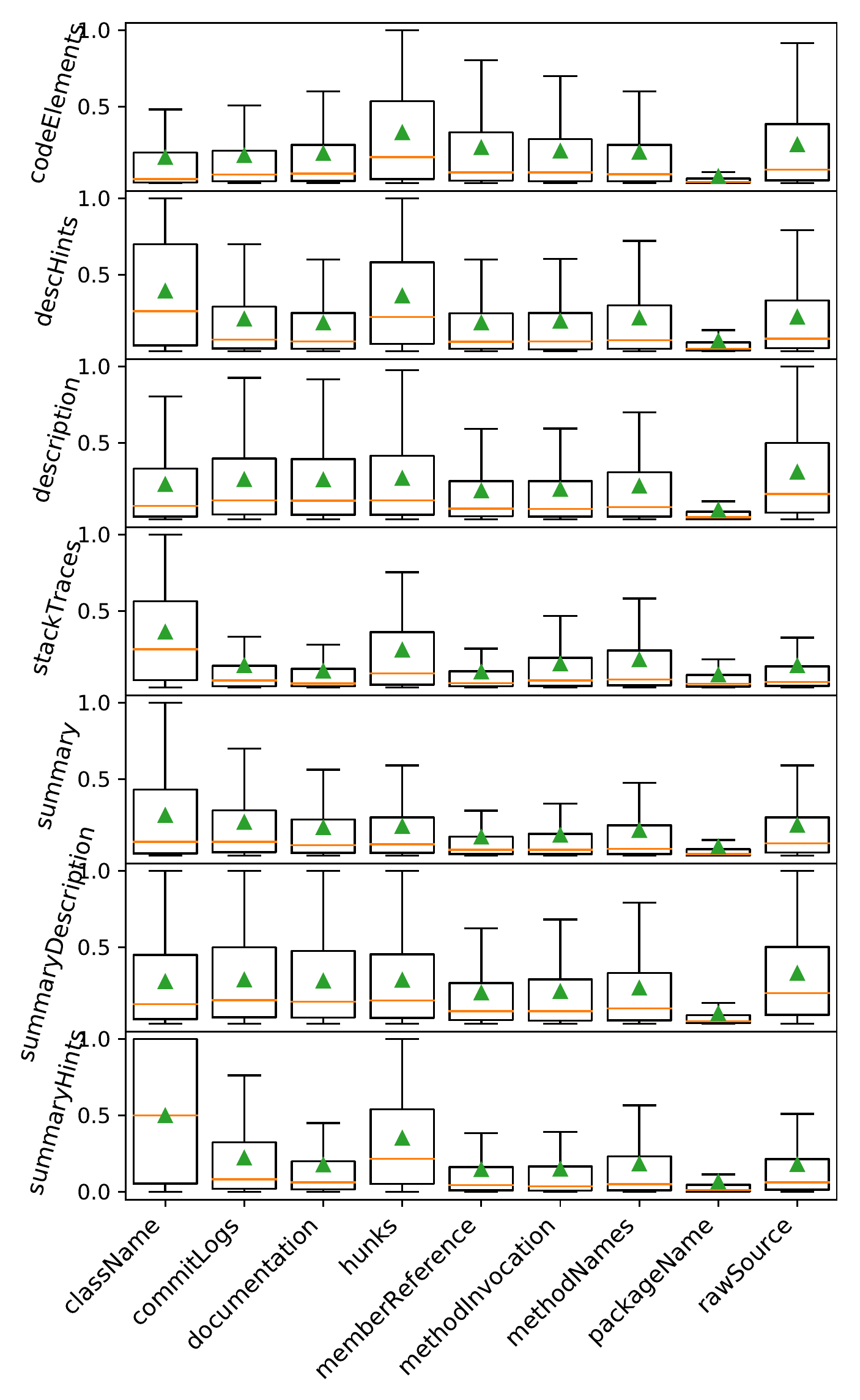}
	\label{fig:singleBB1}
\endminipage\hfill
\minipage[t]{0.3\textwidth}
	\subcaption*{The set of bug reports which are only correctly localized \underline{\em only by BugLocator}, and the location is recommended at Top1 of its list.}
		\includegraphics[width=1\linewidth]{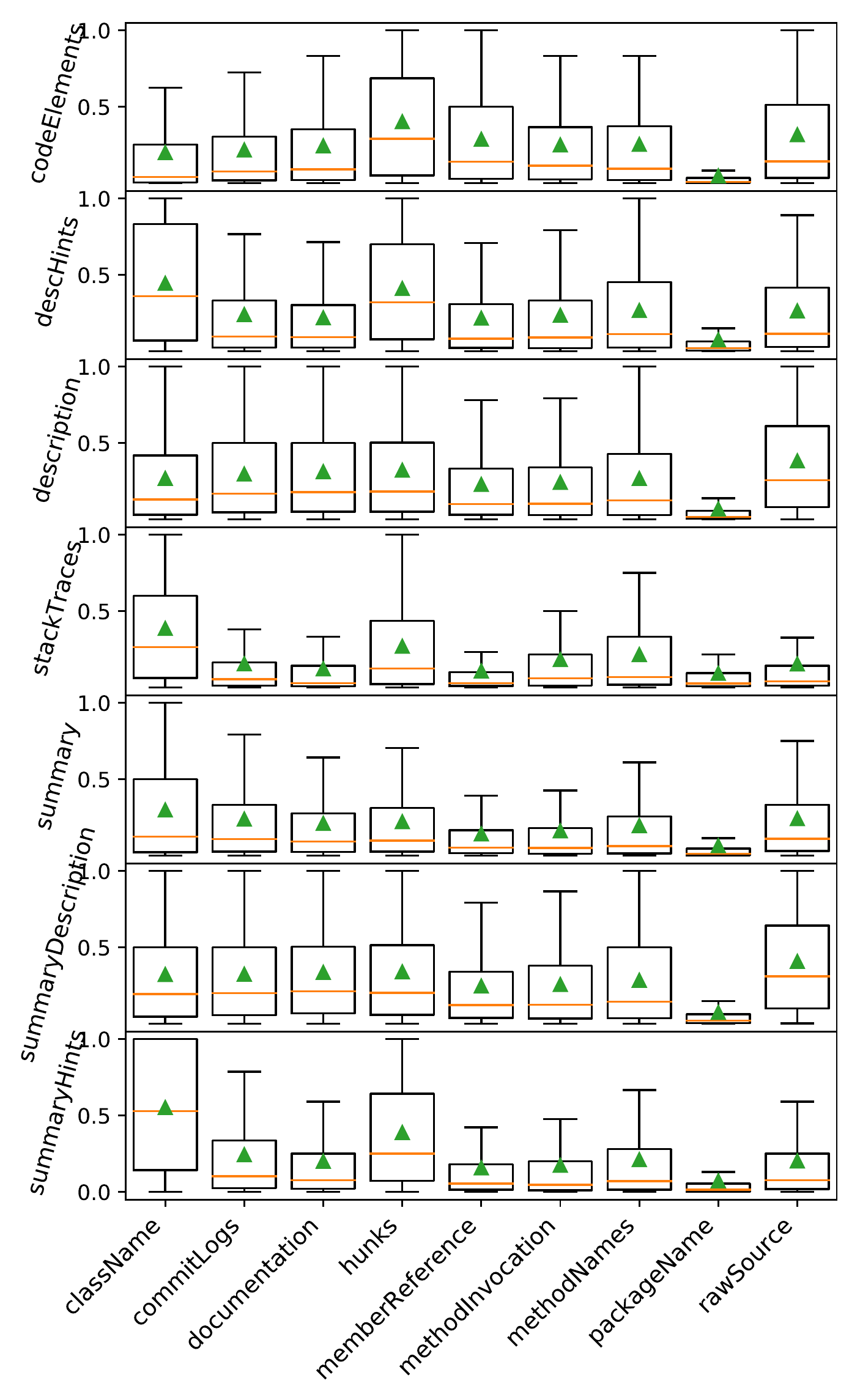}
	\label{fig:singleBB20}
\endminipage\hfill
\minipage[t]{0.3\textwidth}
	\subcaption*{All bug reports for which \underline{\em all} state-of-the-art tools can recommend the correct bug location at Top1 of its list.}
	\vspace{4mm}
		\includegraphics[width=1\linewidth]{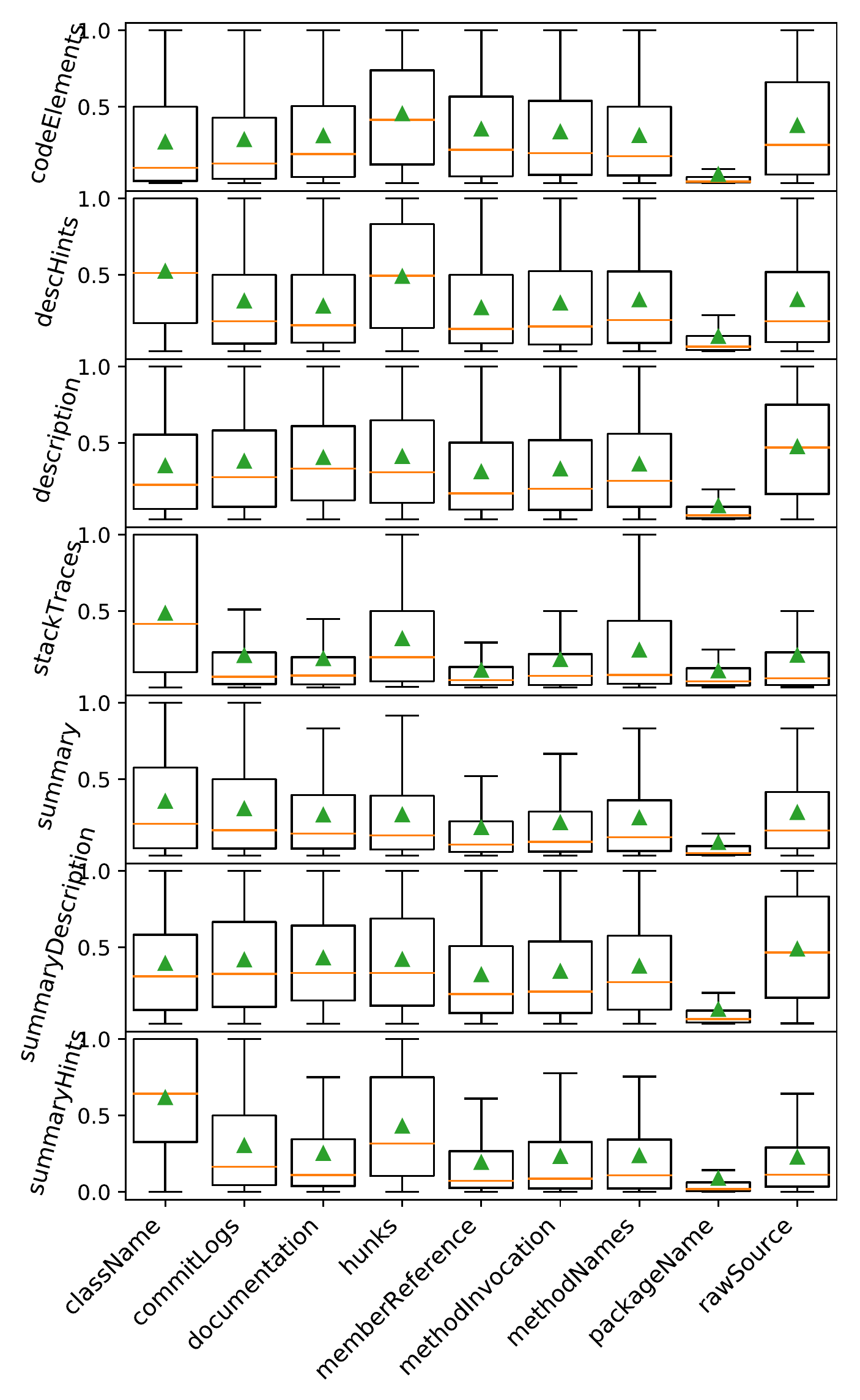}
	\label{fig:singleBB20}
\endminipage\hfill

\vspace{-1cm}
	\caption{Mean Average Precision distributions for different sets of bug reports and with various query formulations (i.e., different combinations of features where the vertical axis refers to the bug report features and the horizontal axis to the source code features.The orange lines show the median values and the green arrows the mean values of the distributions).}
\label{fig:mapDissection}
\end{figure*}

\begin{figure*}

\minipage[t]{0.3\textwidth}
	\subcaption*{All bug reports for which \underline{\em at least one} state-of-the-art tool can recommend the correct bug location at Top1 of its list.}
		\includegraphics[width=1\linewidth]{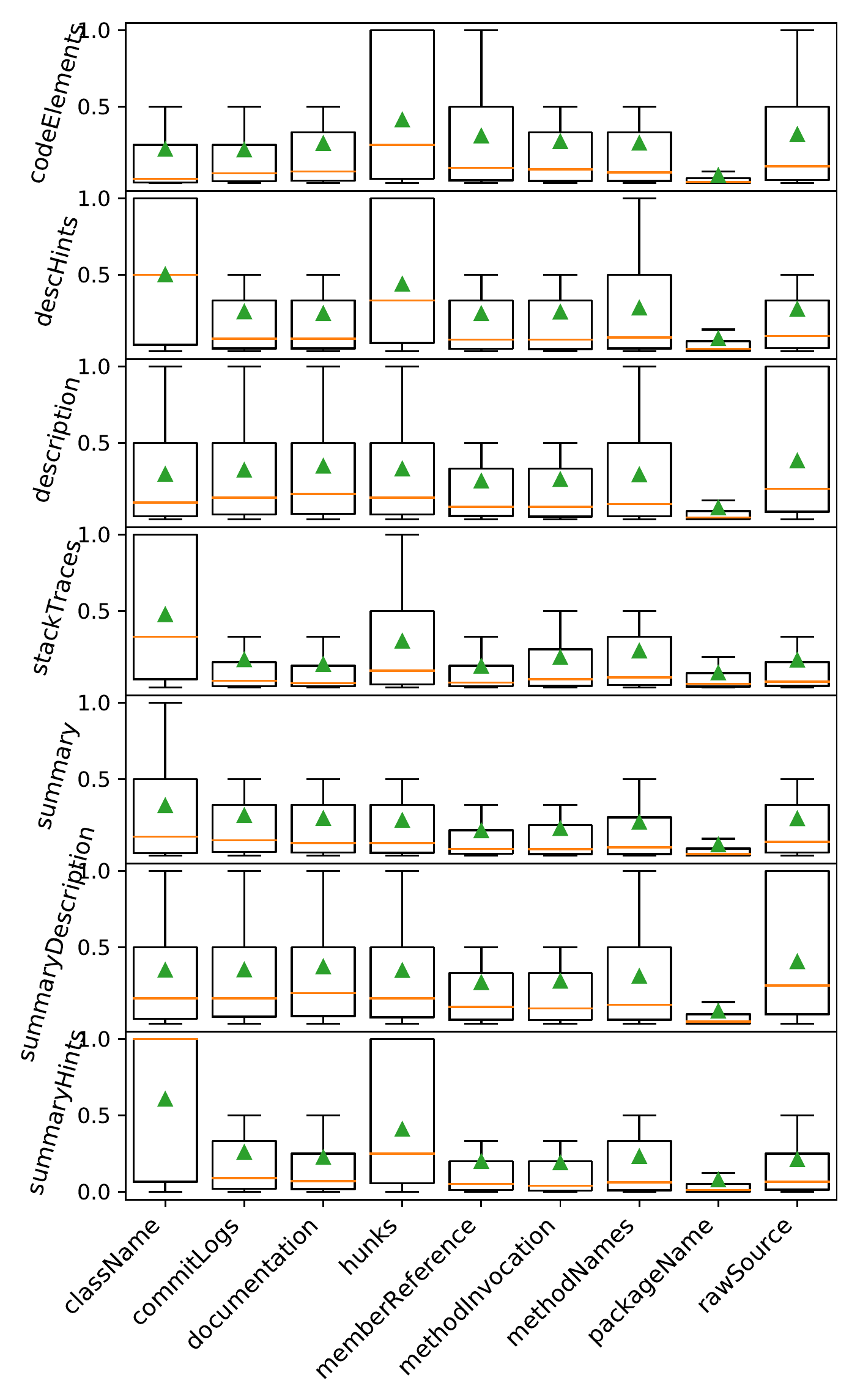}
	\label{fig:singleBB1}
\endminipage\hfill
\minipage[t]{0.3\textwidth}
	\subcaption*{The set of bug reports which are only correctly localized \underline{\em only by BugLocator}, and the location is recommended at Top1 of its list.}
		\includegraphics[width=1\linewidth]{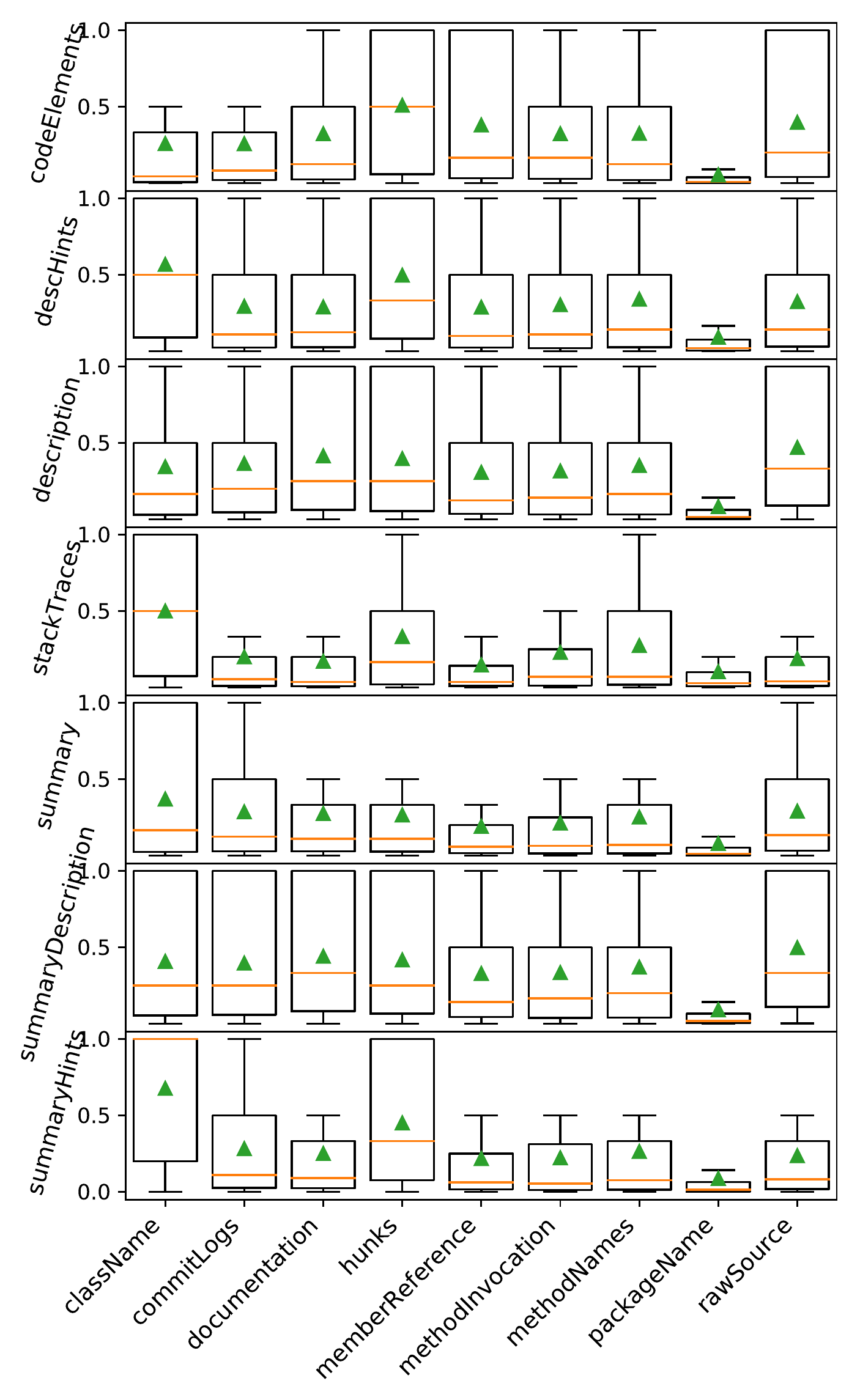}
	\label{fig:singleBB20}
\endminipage\hfill
\minipage[t]{0.3\textwidth}
	
	\subcaption*{All bug reports for which \underline{\em all} state-of-the-art tools can recommend the correct bug location at Top1 of its list.}
	\vspace{4mm}
		\includegraphics[width=1\linewidth]{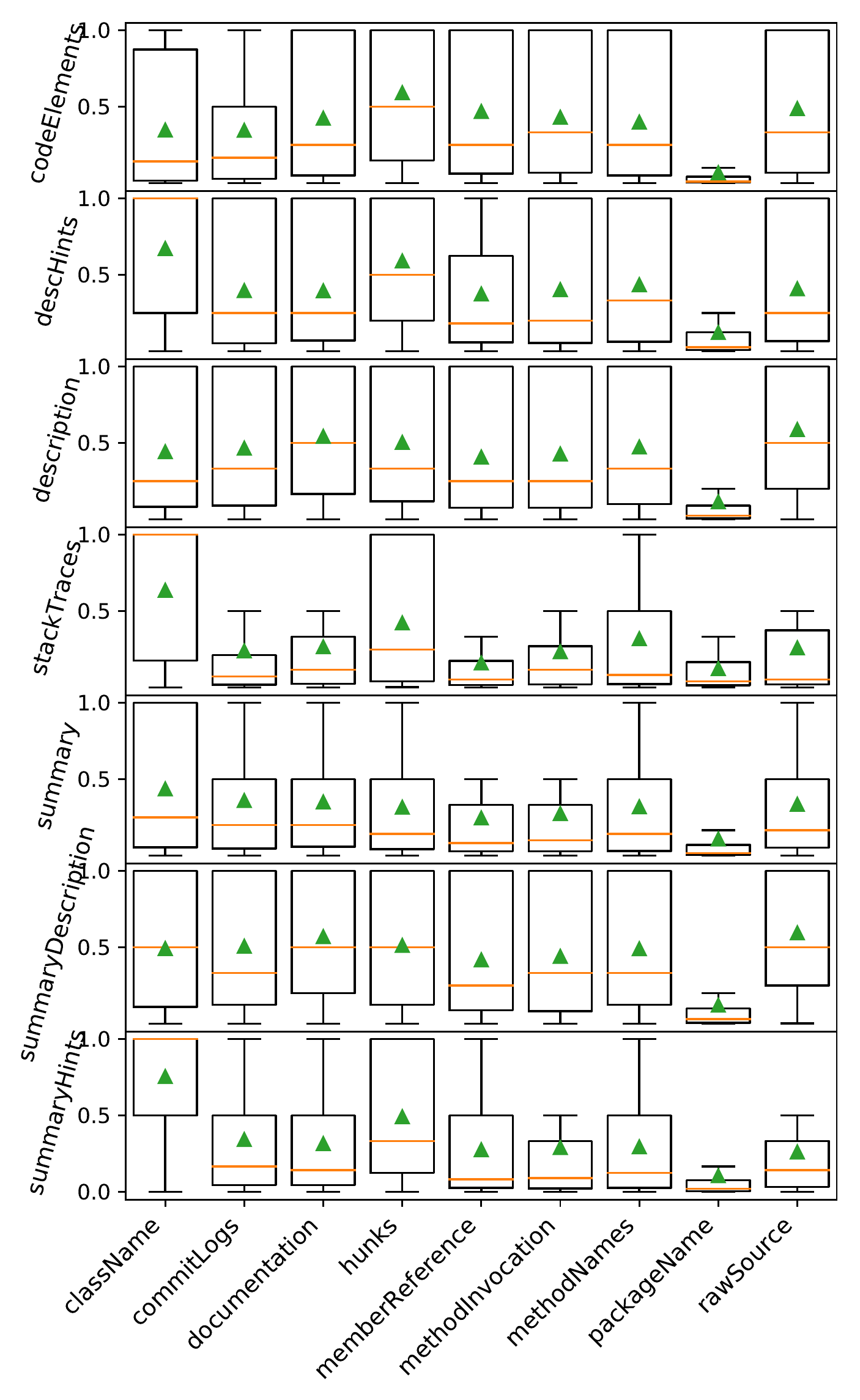}
	\label{fig:singleBB20}
\endminipage\hfill

\vspace{-1cm}
\caption{Mean Reciprocal Rank distributions for different sets of bug reports and with various query formulations (i.e., different combinations of features).}
\label{fig:mrrDissection}
\end{figure*}

Bug localization is then performed by assessing the strength of the similarity in a query-document pair $(q,d)$(cf. Equation~\ref{eq:simi}).  Concretely, we compute similarity matrices for each bug report and all source code files, by considering the pairwise combinations of features (cf. Tables~\ref{tab:bugFeatures} and \ref{tab:codeFeatures}) that are used to represent queries and documents.

\subsubsection{RQ-2: Feature Importance}
To answer our second research question (RQ2), we investigate first 
 the MAP and MRR performance that can be achieved for each pairwise combination 
 of features from bug reports and source code files to highlight the contribution of the features to MAP and MRR performance. Additionally, we perform an  a-posteriori analysis for investigating the discriminating feature combinations that are effective for bug reports which are exclusively localized by specific IRBL tools. The objective of the a-posteriori analysis is to clarify our intuition that IRBL tools are most successful on specific sets of bug reports, which are correlated with the features used for similarity computation.
 
\subsubsection{Results}
We investigate the distribution of MAP and MRR across various bug report sets as illustrated in Figures~\ref{fig:mapDissection} and \ref{fig:mrrDissection}.
 Given a bug report, we produce the ranked list of files to be recommended based on the similarity matrices: the higher the similarity, the higher the localization rank.
Our experiments are performed for different sets of bug reports that we regroup based on the performance of state-of-the-art tools to localize the reported bugs within project source code files.
However, due to space limitation, we only present in this paper three examples sets which are sufficient for building the insights for our subsequent approach:
\begin{itemize}[leftmargin=*]
	\item the set of bug reports that are {\em eventually localizable}, i.e., at least one state-of-the-art tool can place the correct localization file as its Top1 recommendation.
	\item the set of bug reports which appear to be {\em suitable for IR-based bug localization regardless of the tool}. These are then obtained by considered all bug reports for which all state-of-the-art Top1 recommended file was correct.
	\item the set of bug reports which are {\em only localizable by a single tool}. In this case, we focused on bug reports for which only BugLocator can provide a correct Top1 recommended source code file.
\end{itemize}

We note from the distributions of MAP and MRR values in Figures~\ref{fig:mapDissection} and \ref{fig:mrrDissection} that the range of distributions varies across the pairwise combinations of bug report/source code features.
However, in most cases, the median values remain below 0.10 for the MAP as well as MRR. Nevertheless, it is noteworthy that a few combinations stand out by providing significantly high performance.
This dissection illustrates how matching class names, hunks, raw code tokens from source code with code related tokens of bug reports is most effective for bug localization. Indeed, {\tt summaryHints/classNames}, {\tt codeElements/hunks}, {\tt summaryDescription/rawSource} appear to be the most successful pairs in terms of MAP and MRR.


\find{
Information tokens available in bug reports and source code files contribute with varying significance to the performance of bug localization. Only a limited number of pairwise combinations among IR features yield relevant similarity scores for bug localization.
}

Finally, we observe that the different groups of bug reports that have been successfully localized by \underline{all} or \underline{at least one} or \underline{only one} state-of-the-art tool show radically different MAP and MRR distributions for the various pairwise comparisons of features. This finding suggests that the considered sets of bug reports and/or the source code localization files have distinctive attributes which make specific information tokens more or less important for similarity computation.

\find{
Although only a few IR features appear to be effective, it is noteworthy that IR feature pairwise combinations (between bug reports and code files) are not equally significant, in terms of contribution to the localization performance, across the dataset.
}

Figures~\ref{fig:mapDissection} and \ref{fig:mrrDissection} included  detailed MAP and MRR for the dataset of bug reports localized exclusively by BugLocator. We have further performed an extensive a-posteriori analysis of feature importance for different sets bug reports that are exclusively localized by BugLocator, Bluir, Amalgam, BrTracer, Blia and Locus. We perform a Principal Component Analysis in which we rely on a LightGBM model to automatically compute feature importance as a classifier is trained to fit with the bug reports in specific sets. For each set of bug reports exclusively localized by a given IRBL tool, the feature importance values are averaged. Table~\ref{tab:featureImportance} provides details on the number of principal components (features) which capture most of the variance (normalized importance), and the main contributors of those principal components (cumulative importance). 

\begin{table}[!h]
\centering
\caption{Results of Principal Component Analysis.}
\resizebox{1.0\linewidth}{!}{
\begin{tabular}{|l|l|r|r|r|}
\toprule
      &Feature &  Importance & \makecell[c]{Normalized\\importance}  &  \makecell[c]{Cumulative\\importance} \\
      \midrule

\multirow{5}{*}{\rotatebox[origin=l]{0}{Buglocator}}

&            summary2commitLogs &      11.000 &                  0.061 &                  0.061  \\
&                 summary2hunks &       9.000 &                  0.050 &                  0.111  \\
&  summaryDescription2className &       8.000 &                  0.044 &                  0.156  \\

      \midrule

\multirow{5}{*}{\rotatebox[origin=l]{0}{Bluir}}

&        summary2documentation &           4.000 &                  0.400 &                  0.400    \\
&        descHints2className &           3.000 &                  0.300 &                  0.700    \\
&       descHints2commitLogs &           1.000 &                  0.100 &                  0.800    \\

      \midrule
\multirow{5}{*}{\rotatebox[origin=l]{0}{Amalgam}}

&          summary2className &      11.000 &                  0.193 &                  0.193  \\
&     codeElements2className &       9.000 &                  0.158 &                  0.351  \\
&        descHints2className &       5.000 &                  0.088 &                  0.439  \\

      \midrule

\multirow{5}{*}{\rotatebox[origin=l]{0}{Brtracer}}

&               summary2commitLogs &      33.000 &                  0.058 &                  0.058  \\
&                  descHints2hunks &      33.000 &                  0.058 &                  0.116  \\
&          summary2memberReference &      32.000 &                  0.056 &                  0.172  \\

      \midrule

\multirow{5}{*}{\rotatebox[origin=l]{0}{Blia}}

&         summary2documentation &      54.000 &                  0.049 &                  0.049  \\
& summaryDescription2commitLogs &      54.000 &                  0.049 &                  0.097  \\
&            summary2commitLogs &      54.000 &                  0.049 &                  0.146  \\

      \midrule

\multirow{5}{*}{\rotatebox[origin=l]{0}{Locus}}

&       summaryDescription2commitLogs &     124.000 &                  0.077 &                  0.077  \\
& summaryDescription2methodInvocation &      97.000 &                  0.060 &                  0.136  \\
&                 descHints2className &      91.000 &                  0.056 &                  0.193  \\

\bottomrule
\end{tabular}
}
\label{tab:featureImportance}
\end{table}

{\em Importance} provides a score that indicates how useful or valuable each feature was in the construction of the boosted decision trees within the model. The more an attribute is used to make key decisions with decision trees, the higher its relative importance.

This importance is calculated explicitly for each attribute in the dataset, allowing attributes to be ranked and compared to each other. It is calculated for a single decision tree by the amount that each attribute split point improves the performance measure, weighted by the number of observations the node is responsible for. 

The feature importance is then averaged across all of the decision trees within the model as normalized importance. 
From the results, it appears that each exclusively-localized set has a different set of top important feature combinations. We can then conclude that these sets of bug reports require specific weighting scheme by the IRBL tool for the used features.

\begin{figure*}[!t]
\resizebox{\linewidth}{!}{%
	\includegraphics[width=0.75\textwidth]{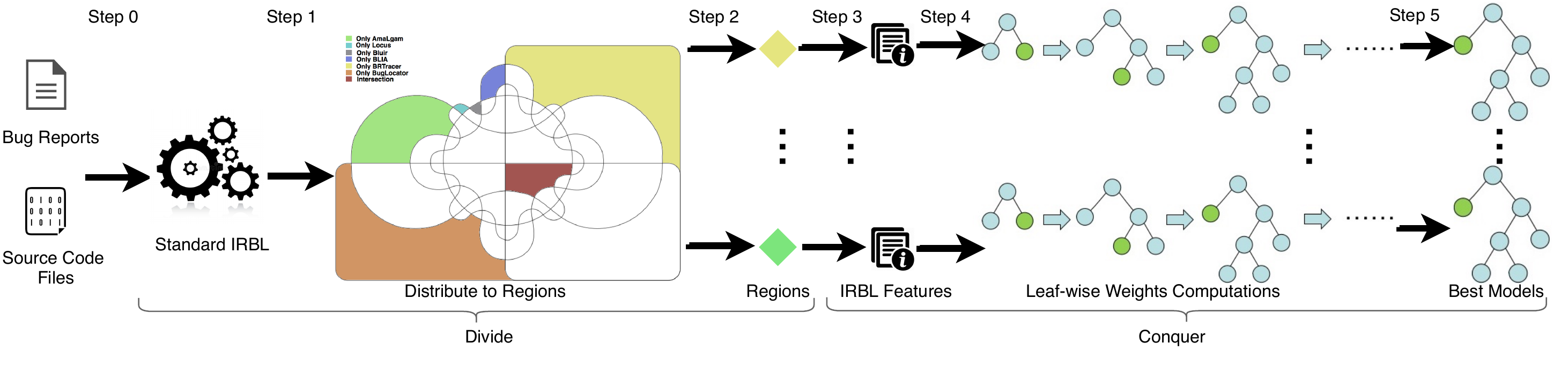}
	}
\caption{\textbf{Divide and Conquer.} Learning Approach.}
\label{fig:approach}
\end{figure*}

\subsubsection{Implications} Findings from the dissection of the impact of IR feature selection on the performance of IRBL suggest two key implications for research:

\begin{enumerate}[leftmargin=*]
	\item {\bf Similarity scores should be weighted}. Since IR features (and their pairwise combinations) have varying significance, the similarity scores that are computed should be weighted accordingly ({\em i.e., explicitly over-rank a bug report/code file pair when a significant feature combination has a relatively high score, and to avoid under-ranking a bug report/code file pair when an otherwise irrelevant feature combination has a low score.}). In the literature of IRBL, a number of tools (see reproducibility study in ~\cite{lee2018bench4bl}) leverage a sampled dataset to compute the scores that guarantees the best localization performance. These weighted scores are then used as generic scores for the overall approach.
	\item {\bf Weights of IR features should be adaptively computed for every specific set of bug reports}. Our experiments showed that relying on the similarity score for a pair of IR features (e.g., {\tt summaryHints} in the bug report and {\tt methodNames} in source code file) can lead to varying performance of IRBL depending on the dataset. It is thus important to weigh these scores differently in accordance with the ``nature'' of the bug report and source code file that are being compared.
\end{enumerate}

\section{D\&C: an Approach to Adaptively Learn the Weights of Similarity Scores}
\label{sec:approach}

The implications enumerated in the previous section provide a motivation for building
a learning approach for adaptively computing the most effective weights to apply to the similarity scores of IR features of a given pair of bug report/source code file. We explore such an approach with a supervised learning technique where a classification is built by learning from the examples of the dataset.  However,
building on the insights of our findings, we consider a multi-classifier approach where several classifiers are built, each being trained on specific parts of the dataset. During classification, instead of selecting a single classifier and using its probability outputs to present a ranked list of localized files, we combine the outputs of all classifiers by averaging the prediction probabilities. 

\subsection{Feature Space for the Classification Models}
A classification model in machine learning-supported IR-based bug localization is trained on a sample dataset to accurately suggest whether a given source code file is a likely location of the reported bug. Generally, this decision is taken by learning the relevant weights for the similarity scores and by computing the appropriate cutoff similarity threshold. In our case, we ignore the learned cutoff and return a ranked list of files based on the probabilities.
In practice, we feed the learning algorithm with feature vector representations of the bug reports and source code files. We directly leverage the similarity matrices presented earlier, leading to 70-dimension feature vectors (with similarity scores of the pairwise combinations between the 7 bug reports information types and the 10 source code information types described in Section~\ref{sec:dissection}).

\subsection{Divide-and-Conquer via Multi-classification}
The first challenge for building a multi-classifier is to identify an effective way of splitting the dataset into {\em meaningful subsets}, i.e., subsets where samples share commonalities with respect to the significance of specific IR features for IRBL.
To that end, an immediate approach would be to build metrics which first cluster the samples according to the distribution of MAP and MRR (based on a baseline IRBL tool) for pairs that provide similar localization performance for the same feature subsets. 
This would require however an exhaustive exploration of combinations beyond the 70 pairwise combinations investigated in this study, as one would then need to investigate various combinations for $K = X + Y$ features, where $X\in [1..7]$ represent the number of bug report features and $Y\in [1..10]$ represent source code features.

Nevertheless, our empirical study of IRBL tools suggests that each tool appears to be more successful than others in some specific regions of the datasets while another large region seems to be localizable similarly by all tools. Given that our thorough review of the recent literature revealed that most state-of-the-art mainly differ by the set of features that are explored for the similarity computation, it would be reasonable to assume that the regions represent sets of bug reports/code files where the computed weights are effective. Thus, we propose to directly leverage the assessment result of state-of-the-art tools as a metric to split the training dataset.

Figure~\ref{fig:approach} overviews the overall divide-and-conquer approach. We start by running and assessing state-of-the-art tools on Bench4BL, and delimitating different regions of the datasets according to the performance of each tool in comparison with others. Then, for each of such regions, we apply the learning process formally detailed in Algorithm~\ref{alg:dc}. The regions contain the list of bug reports ids, thus
	we start by extracting  the fine-grained 7x10 IRBL features ($f$), and their corresponding binary labels ($l$) for the bug report/code files pairs. In a given project, only a few source code files are buggy ($trueLabels$) and most of them are not ($falseLabels$) which is creating a highly imbalanced dataset. To handle this imbalance nature, we calculate a frequency coefficient ($freqCoef$) that represents the ratio of the false labels observed in the training data for each project. Then, we  apply the frequency coefficient over the labels observed in the training data to compute the class weights ($cw$). In the class weight computation we add 1 to the label values, in order to prevent having the class weights of falseLabels all zero. We encode the features, labels and class weights into dataset object ($dtrain$). We retrieve hyperparameters of the training ($p$) that are previously calculated using a grid search approach. We train the region classifier for 10000 iterations, with early stopping of 10 rounds until a best model is found. Finally, we predict the testing data on the best model and save the prediction probabilities.

We consider specific subsets of the dataset for training  since they are presenting the following properties:
\begin{enumerate}[leftmargin=*]
	\item bug reports that fit a specific tool (e.g., {$Brtracer$} represents the subset of bug reports which were accurately localized by  Brtracer).
	\item bug reports are exclusively fit to a specific tool (e.g., {$Only-Brtracer$} represents the subset of bug reports which were accurately localized by only Brtracer).
	\item bug reports that  appear to be rich in terms of information and are thus easy to localize by any tool ({$INTER$}).
	\item bug reports that are localizable at the Top1 position by at least one tool ({$UNION$}).
	\item bug reports that do not seem to contain enough information to be accurately localized by any tool ($\neg{UNION}$).
\end{enumerate}

\begin{algorithm}[!h]

    \SetKwFunction{createDataset}{createDatasetObject}
    \SetKwFunction{KFold}{timelineSplit}
    \SetKwFunction{extractFeaturesAndLabels}{extractFeaturesAndLabels}
    \SetKwFunction{calculateWeights}{calculateClassWeights}
    \SetKwFunction{setTrainingParams}{getHyperparameters}
    \SetKwFunction{tm}{trainModel}
    \SetKwFunction{predict}{best\_model.predict}

    \SetKwInOut{Input}{input}
    \SetKwInOut{Output}{output}
    \SetKw{Return}{return}

    \Input{A region $r$ from regions $R$ which is a list of bug reports,
    where $r \subset{R}$ and
    $R$  $\leftarrow$ \( \{ $\mbox{Only-Brtracer}, $\mbox{Only-LOCUS},
    $\mbox{Only-Blia}, $\mbox{Only-Amalgam}, $\mbox{Only-Bluir}, $\mbox{Only-BugLocator},\)
    Intersection\(, $\mbox{Top-1s}, $\mbox{Non-Top1s} \} \)}

    \Input{A region $\neg{r}$,
    where  $ \neg{r} ={R \setminus r}$ }
    \Output{Predicted probabilities}

    \For{$r$ \KwTo $R$}{

        $trains$,$tests$ $\leftarrow$ \KFold{$r$}\;
         \For{$train$,$test$ in $trains$,$tests$}{

            $f$,$l$ $\leftarrow$ \extractFeaturesAndLabels{$train$}\;

            $cw$ $\leftarrow$ \calculateWeights{$l$}\;
            $dtrain$ $\leftarrow$ \createDataset{$f$,$l$,$cw$}\;
            $p$ $\leftarrow$ \setTrainingParams{}\;

            $best\_model$ $\leftarrow$ \tm{$dtrain$,$p$}\;
            $prediction\_probabilities$ $\leftarrow$ \predict{$test$}\;
            \Return $prediction\_probabilities$
        }
    }
    \SetKwProg{Fn}{Function}{}{end}
    \SetKwFunction{split}{shuffleAndSplitData}
    \SetKwFunction{filter}{filter}
    \SetKwFunction{cou}{count}

    \Fn{\calculateWeights(l: labels) : classWeights}{
     $trueLabels$ $\leftarrow$ \filter{$l$,$1$}\;
     $falseLabels$ $\leftarrow$ \filter{$l$,$0$}\;
     $freqCoef$ $\leftarrow$ {\cou{$falseLabels$}} / {\cou{$trueLabels$}}\;
     
     $classWeights$ $\leftarrow$ (l  * $freqCoef$) + 1 \;
    \Return $classWeights$\;}

    \caption{Divide and Conquer Learning Algorithm}
    \label{alg:dc}
\end{algorithm}


\noindent{\textbf{Splitting the dataset into test and training}.}
For each selected region $r$, we first prepare the dataset for validation by splitting it into validation and training data based on the bug report creation year (i.e., $timelineSplit$ in Algorithm~\ref{alg:dc}): we train on bug reports with creation date less than X and validate on creation date greater than X, where X $\in \{1-1-2008, 1-1-2009,..., 1-1-2016\}$.

	
\noindent{\textbf{Handling dataset imbalance.}}	
	 The whole dataset is highly imbalanced since only a very small portion ($\approx$ 20\,000 out of 35\,088\,421) of the bug report-source code pairs are actually buggy ({\em i.e. minority class}), whereas majority of the pairs are non-buggy ({\em i.e. majority class}). Machine learning classifiers have a bias towards classes which have a large number of instances and thus tend to only predict the majority class data. The features of the minority class are then treated as noise and are often ignored. Therefore, there is a high probability of misclassification of the minority class as compared to the majority class, and hence the likelihood of poor classification for bug localization.
	 Dealing with imbalanced datasets entails strategies such as either improving classification algorithms, or balancing classes in the training data before providing the data as input to the machine learning algorithm.
	 We initially experiment on balancing the classes in the training dataset by applying resampling techniques (Random Under-Sampling, Random Over-Sampling, Cluster-Based Over Sampling~\cite{yen2009cluster}, SMOTE~\cite{chawla2002smote}, MSMOTE~\cite{hu2009msmote}). The main objective of these techniques is to {\em artificially} balance classes by either increasing the frequency of the minority class or decreasing the frequency of the majority class. However, we note that resampling techniques are not very effective since these random resampling techniques are {\em overfitting} the training data, and others (Cluster-Based Over Sampling~\cite{yen2009cluster}, SMOTE~\cite{chawla2002smote}, MSMOTE~\cite{hu2009msmote}) are not performing well due to the high dimensionality of data (70-dimension feature vectors).
	  
	 We propose to address the imbalance dataset issue by adapting the classification algorithm. Our machine learning classification is performed with the LightGBM gradient boosting framework~\cite{ke2017lightgbm}. LightGBM documentation\footnote{\url{https://lightgbm.readthedocs.io/en/latest/Parameters.html}} provides parameters to explicitly instruct the algorithm to account for data imbalance. It is further known in the practice of machine learning (e.g., Kaggle competitions) to be effective for dealing with imbalanced dataset\footnote{\url{https://www.kaggle.com/pranav84/lightgbm-fixing-unbalanced-data-lb-0-9680}}. LightGBM supports weighted training, which uses the observation weights of classes for bias correction. We compute the class weights as described in Algorithm~\ref{alg:dc}. We filter the training data by selecting positive and negative samples where the positive samples have the label 1, which presents the actually buggy bug report-source code pairs ({\em i.e., minority class}), and negative samples have the label 0, that presents the non-buggy pairs ({\em majority class}). We calculate the frequency coefficient, which is a parameter that is inversely proportional to class frequencies in the training data for each project. Finally, we calculate the class weights by applying the frequency coefficient over the labels observed in the training data. 
	 During training, LightGBM which uses a Leaf-wise (Best-first) Tree Growth approach will choose the leaf with max delta loss to grow. Due to the larger loss function pre-factor ({\em i.e., class weights}) the classes with higher weights matter more, even they are minority class.
	 

%

\noindent{\textbf{Hyperparameter optimization}.}	 
LightGBM uses the leaf-wise tree growth algorithm, while many other popular tools use depth-wise tree growth. Compared with depth-wise growth, the leaf-wise algorithm can converge much faster. However, the leaf-wise growth may be over-fitting if not used with the appropriate parameters. To build optimal models using a leaf-wise tree growth approach, there are a few important parameters to tunes:
\begin{itemize}[leftmargin=*]
\item Number of estimators. The number of trees in the model.
\item Number of leaves. This is the main parameter to control the complexity of the tree model. Theoretically, we can set $num\_leaves = 2^{(max\_depth)}$ to obtain the same number of leaves as a depth-wise tree. However, this simple conversion is not good in practice. The reason is that a leaf-wise tree is typically much deeper than a depth-wise tree for a fixed number of leaves. Unconstrained depth can induce over-fitting. Thus, when trying to tune the $num\_leaves$ parameter, we should let it be smaller than $2^{(max\_depth)}$. 
	
\item Learning rate. This a hyper-parameter that controls how much we are adjusting the weights of our network with respect to the loss gradient.

\item Feature fraction. The fraction of observations to be selected for each tree. Selection is done by random sampling.
\end{itemize}

We used a grid search which systematically works through multiple combinations of parameter tunes, cross validating all runs to determine which one gives the best performance. Eventually, for the training, a binary classifier is constructed with $learning\_rate = .03$, $n\_estimators = 100$, $num\_leaves = 31 $ and $feature\_fraction = .08$.

For runtime efficiency, we have transformed our training data into LightGBMs' inbuilt dataset object where the features, labels, and weights are represented in a memory efficient binary form. 
	
\noindent{\textbf{Models Selection and Prediction.}} We train each classifier in 10\,000 iterations, with {\em early stopping}, which is a form of regularization used to avoid overfitting when training a learner with an iterative method. 
Early stopping works by monitoring the performance of the model that is being trained on a separate test dataset and stopping the training procedure once the performance of the test dataset has not improved after a fixed number of training iterations. It avoids overfitting by attempting to automatically select the inflection point where performance on the test dataset starts to decrease while performance on the training dataset continues to improve as the model starts to overfit.
Concretely, we can check whether there is no improvement via the root mean squared error (RMSE) over the 10 consecutive iterations during the training. Once there is no improvement, the training early stops since the $best\_model$ is found ( or in the worst case scenario the iteration count reaches to 10\,000 which terminates the training by selecting the last iteration as $best\_model$).

Once the $best\_model$ for each classification training is found, we apply it to the relevant test dataset, yielding probability values for all pairwise combinations of bug report-source code files. A high probability value implies that the classifier highly recommends the source code file in a pair to be relevant to the associated bug report.
	
\subsection{Ranking of Bug Localization Recommendations}
The prediction probabilities of the source code files in each model are combined into a single ranking by averaging the prediction probabilities. We use a classical incremental ranking, ties are ignored and the ranks are assigned in the order they appear. For example, when files A and B have the same prediction probability, the classical incremental ranking ranks A in rank $r$, and B in rank $r+1$.

\section{Assessment}
\label{sec:evaluation}
The assessment of this work investigates execution times and performance results in validation experiments.
We also perform a comparison against the state-of-the-art and study the impact of the choice of multi-classification.

\subsection{Execution Times}

The training of 140 classifiers and all ($\sim70\,000$) predictions took 69,25 hours on a server with 110 Intel Xeon E5-4650v2@2.4GHz with 4TB of RAM.

The feature extraction processes are executed separately in parallel for every project. The execution speed is roughly proportional to the number of source code files in the project and the number of bug reports. Most of the small projects (projects having less than a million bug report-source code pairs) finish within an hour, while the bigger projects JBOSS-WFLY, APACHE-HIVE took around 50 hours to complete the feature extraction. For the biggest project APACHE-CAMEL, we executed the feature extraction process (including parsing) more than 80 hours on our 96 Intel Xeon E5-4650v2@2.4GHz with 4TB of ram.
 
 \subsection{Validation experiment}
 To assess the \toolname approach, we propose a validation experiment where dataset is split around per year.
 Bug reports (and associated localization information) created before 1 January of the selected year are considered as training data. All bug reports from 1 January and onwards are used for testing (i.e., validation) of the model.  Table~\ref{tab:validationExp} shows the validation experiment results for each year. 
 Overall, the training dataset includes 5321 bug reports, and the built localizer is tested on 3954 bug reports. We record an MAP of 0.507 and an MRR of 0.617. It should be noted that this performance is obtained with a cleaned dataset (i.e., where considered bug reports for validation are not post-fixing activities).
  \begin{table}[!h]
\centering
		\scriptsize

\caption{Validation Experiment.}%
\label{tab:validationExp}

\resizebox{\linewidth}{!}{%
\begin{tabular}{l|l|r|r|r|r|r|r|r|r|r|c}

 &\# bug reports & 2008 & 2009 & 2010 & 2011 & 2012 & 2013 & 2014 & 2015 & 2016 & Average \\
 \cline{3-11}
  &  in training & 79 & 368 & 825 & 1293 & 1619 & 1936 & 2328 & 3102 & 4213 & Perf.\\
  &  in validation & 220 & 345 & 366 & 262 & 251 & 316 & 565 & 834 & 795 & (mean)\\
   \midrule
 & MAP & 0.504 & 0.443 & 0.511 & 0.589 & 0.553& 0.562& 0.495& 0.494& 0.494& {\bf 0.52}\\
 & MRR &  0.635 & 0.571 & 0.624 & 0.693 & 0.646 & 0.678& 0.603& 0.602& 0.596& {\bf 0.63}\\
 \midrule
 \multirow{3}{*}{\rotatebox[origin=l]{90}{\% localized }} \\
   & Top1 & 52\% & 43\% & 50\% &  56\% & 52\%& 57\%& 49\%& 48\% & 47\% & {\bf 50\%}\\
   & Top5 & 75\% & 75\%  & 77\% & 84\% & 79\%&  80\%& 74\%& 75\% & 74\% & {\bf 77\%}\\
   & Top10 & 84\% & 85\%  & 85\%  & 90\% & 86\%&  87\%& 80\%& 83\% & 83\% & {\bf 85\%}\\

\end{tabular}
}
\end{table}
  
  On average, we find that 50\% of bugs can be localized at Top1 by \toolname, and 77\% of bugs at Top5 while 85\% of bugs are detected at Top10.

 \subsection{Comparison against the state-of-the-art} 
To compare \toolname against the state-of-the-art, we consider the execution results of IRBL tools compiled by Bench4BL and focus on results of bug reports that are part of our cleaned validation set (i.e. those bug reports that are truly pre-fix / fix-independent reports). Given that Bench4BL only provides information about bug reports in Top10 for each tool, our comparisons are based, for each tool, on a different set. Table~\ref{tab:perStateOfARt} provides performance comparison details about MRR and MAP. We note that \toolname provides a substantial performance improvement of MAP and MRR over all tools: MAP is improved by between 4 and up to 10 percentage points, while MRR is improved by between 1 and up to 12 percentage points. Furthermore, we note that \toolname generally manages to localize more bug reports at Top1 (e.g., 20\% more than Blia), Top5 (e.g., 15\% more than Bluir) and Top10 (e.g., 13\% more than Bluir).

\begin{table}[!h]
\caption{Performance comparison against state-of-the-art IRBL tools. Dataset are cleaned to fit our criteria on pre-fix activities.}
\label{tab:perStateOfARt}
	\begin{tabular}{|l|r|r|r|r|r|r|}
		\toprule
				  \multicolumn{4}{|r}{Performance} & \multicolumn{3}{|r|}{\% Localized Bug Reports}  \\
		\midrule
		Tool		&\makecell[c]{\# Bug \\Reports} & MAP &MRR & Top1 & Top5 &Top10 \\
		\midrule
  BugLocator & 3949 & 0.425 & 0.546 & 42\% & 69\% & 78\%\\
    D\&C     & 3949 & 0.507 & 0.617 & 50\% & 76\% & 84\%\\
       \midrule 
           
    Brtracer & 3949 & 0.461 & 0.599 & 49\% & 74\% & 81\%\\
    D\&C     & 3949 & 0.507 & 0.617 & 50\% & 76\% & 84\%\\
       \midrule
       Bluir & 3947 & 0.402 & 0.493 & 37\% & 64\% & 74\%\\
    D\&C     & 3947 & 0.507 & 0.617 & 50\% & 76\% & 84\%\\
       \midrule
     Amalgam & 3952 & 0.405 & 0.496 & 37\% & 65\% & 74\%\\
    D\&C     & 3952 & 0.508 & 0.617 & 50\% & 76\% & 84\%\\
          \midrule
        Blia & 3879 & 0.434 & 0.554 & 43\% & 71\% & 80\%\\
    D\&C     & 3879 & 0.514 & 0.625 & 50\% & 77\% & 85\%\\
           \midrule
       Locus & 3362 & 0.422 & 0.604 & 49\% & 75\% & 83\%\\
    D\&C     & 3362 & 0.506 & 0.618 & 50\% & 76\% & 85\%\\
       
 \bottomrule

	\end{tabular}
\end{table}

Table~\ref{tab:perStateOfARt} further details the detection performance of the tools in terms of percentage of localized bug reports at Top1, Top5 and Top10. \toolname localizes up to 13\% more bugs at Top1 than the state-of-the-art tools, between 1\% and up to 11\% more bugs at Top5.

\begin{figure*}[h!]
		\caption*{Mean Average Precision comparisons for the 45 Projects}
\minipage[t]{0.165\textwidth}
		\includegraphics[width=1\linewidth]{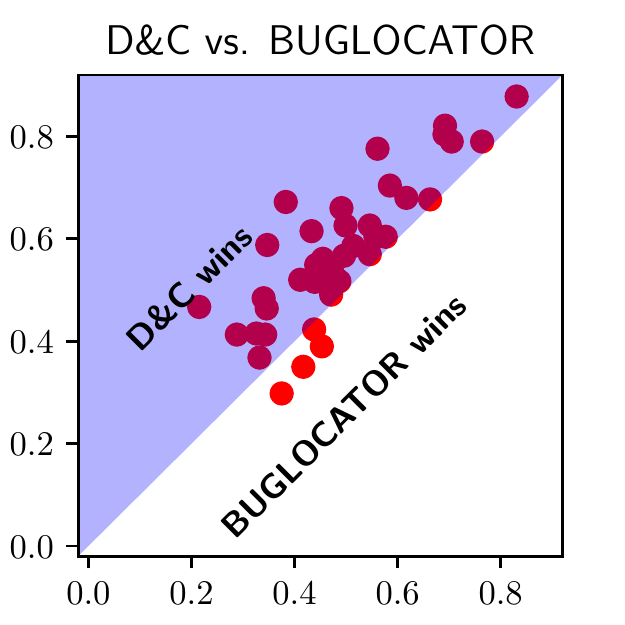}
\endminipage\hfill
\minipage[t]{0.165\textwidth}
		\includegraphics[width=1\linewidth]{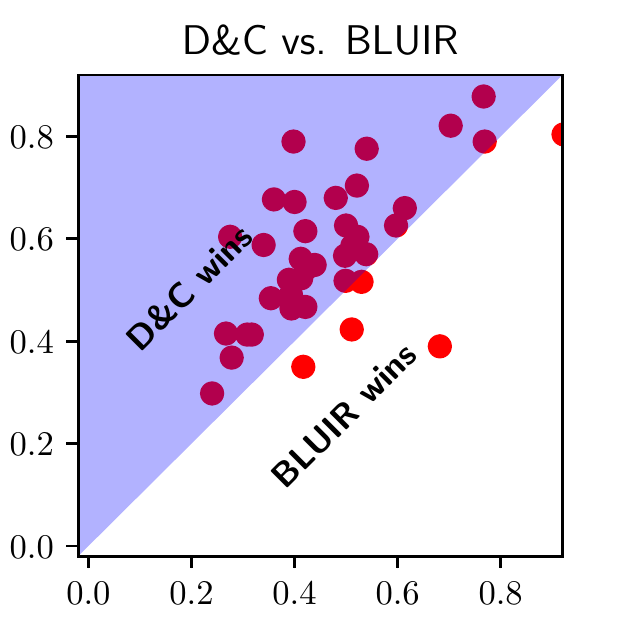}
\endminipage\hfill
\minipage[t]{0.165\textwidth}
		\includegraphics[width=1\linewidth]{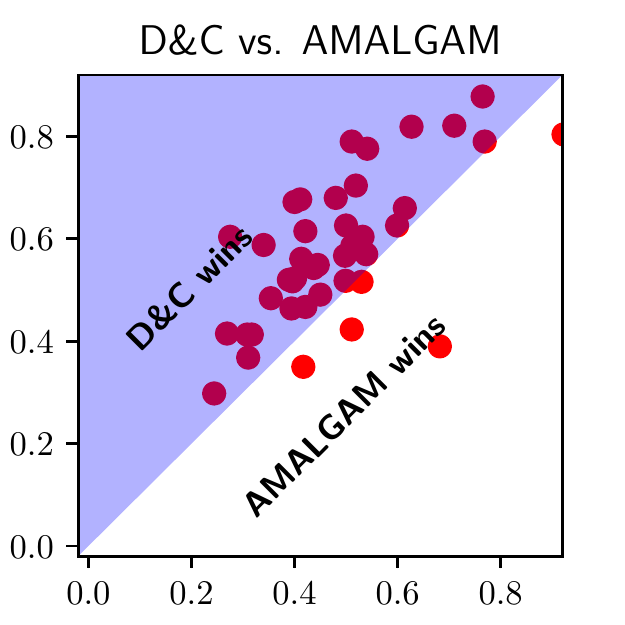}
\endminipage\hfill
\minipage[t]{0.165\textwidth}
		\includegraphics[width=1\linewidth]{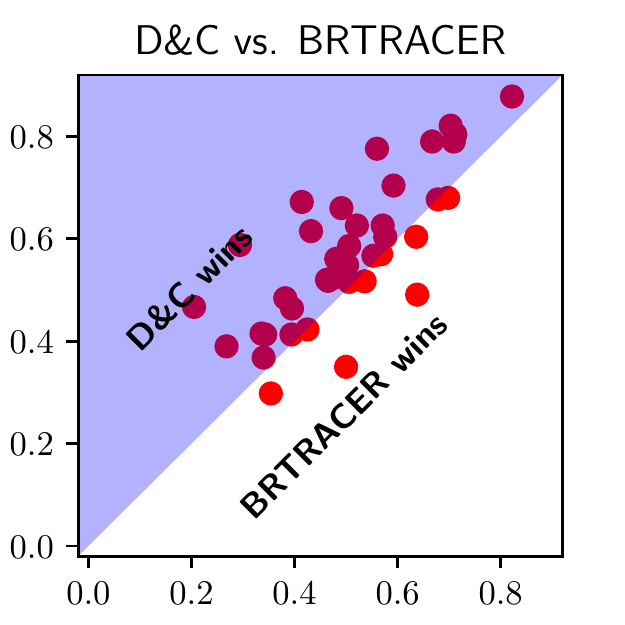}
\endminipage\hfill
\minipage[t]{0.165\textwidth}
		\includegraphics[width=1\linewidth]{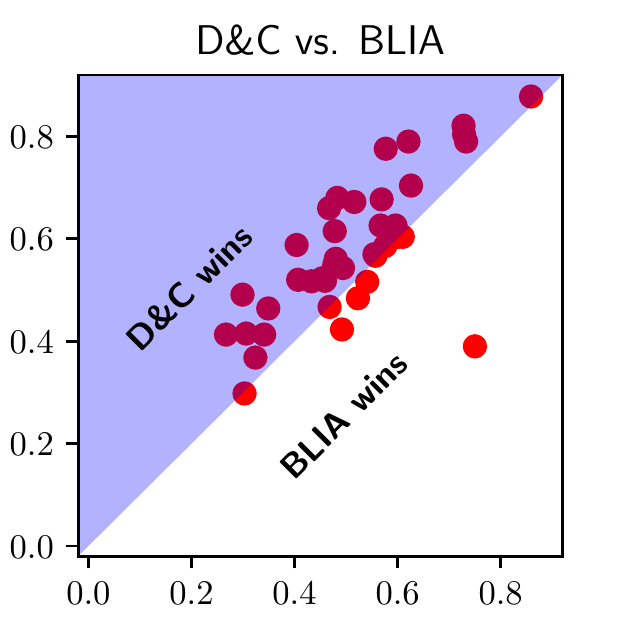}
\endminipage\hfill
\minipage[t]{0.165\textwidth}
		\includegraphics[width=1\linewidth]{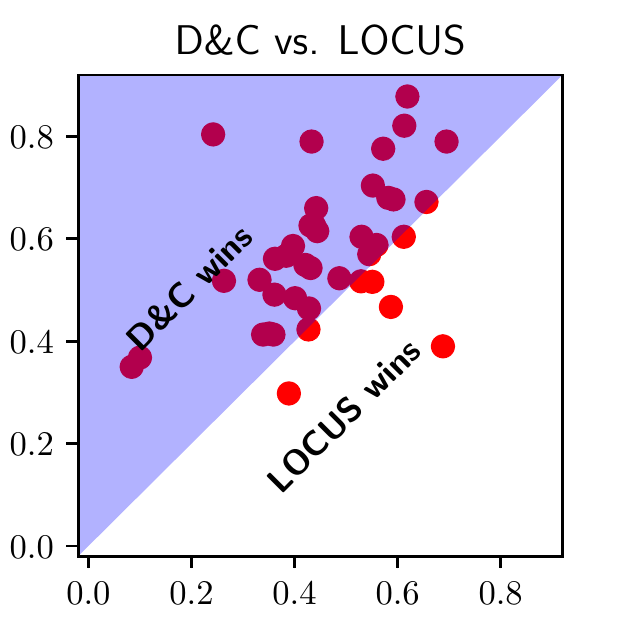}
\endminipage\hfill
		\caption*{Mean Reciprocal Rank comparisons for the 45 Projects}
\minipage[t]{0.165\textwidth}
		\includegraphics[width=1\linewidth]{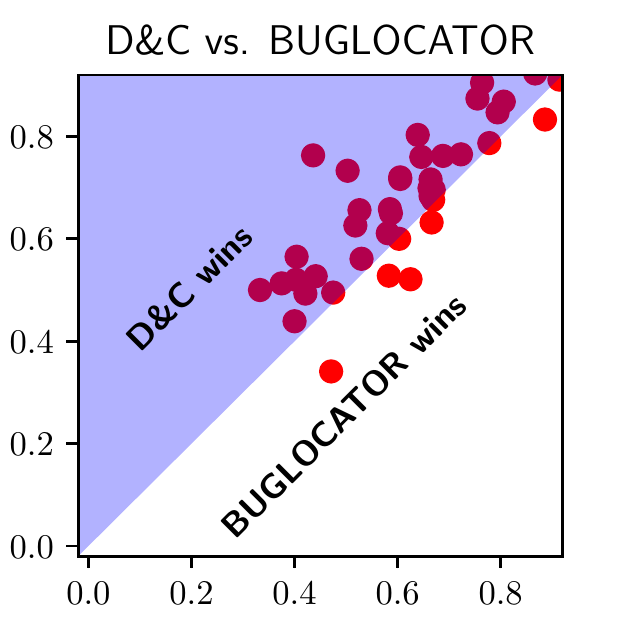}
\endminipage\hfill
\minipage[t]{0.165\textwidth}
		\includegraphics[width=1\linewidth]{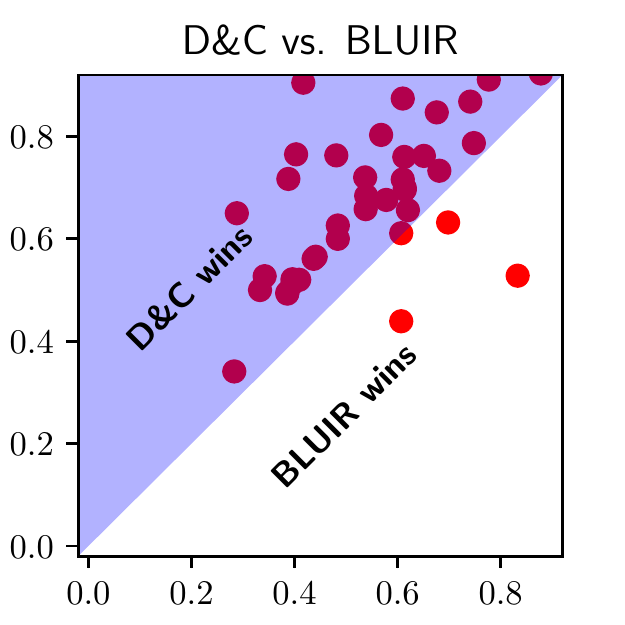}
\endminipage\hfill
\minipage[t]{0.165\textwidth}
		\includegraphics[width=1\linewidth]{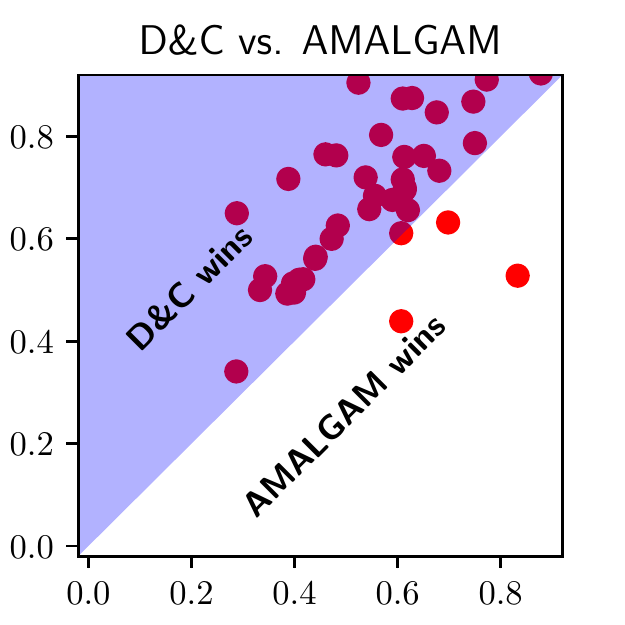}
\endminipage\hfill
\minipage[t]{0.165\textwidth}
		\includegraphics[width=1\linewidth]{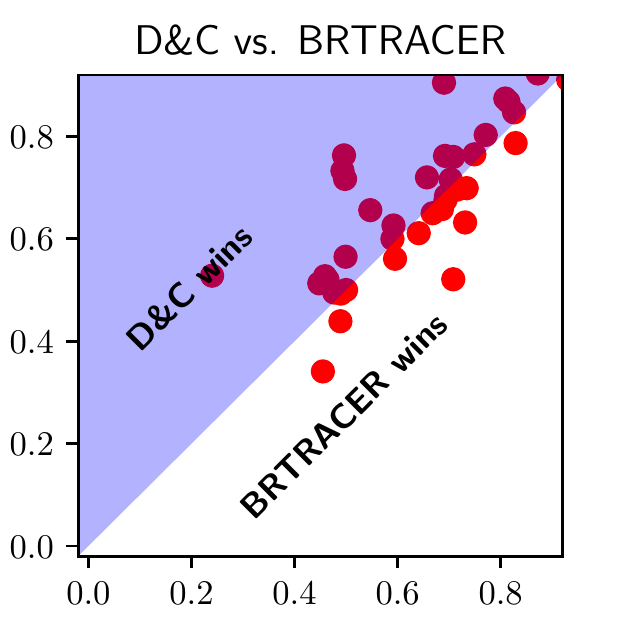}
\endminipage\hfill
\minipage[t]{0.165\textwidth}
		\includegraphics[width=1\linewidth]{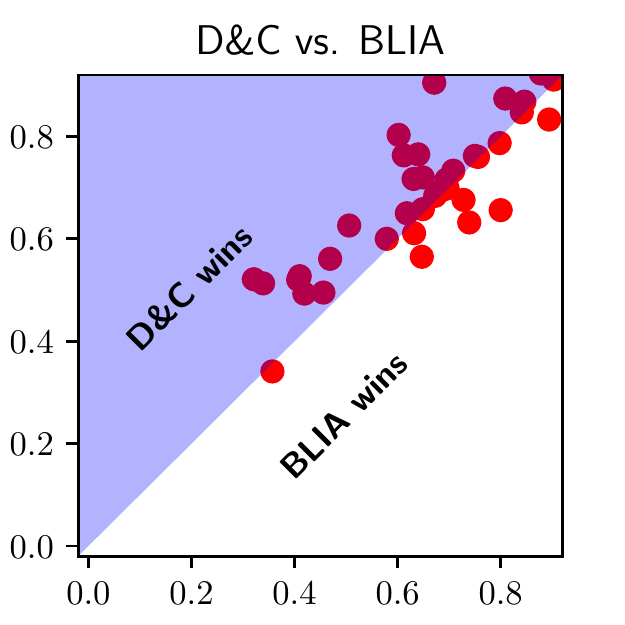}
\endminipage\hfill
\minipage[t]{0.165\textwidth}
		\includegraphics[width=1\linewidth]{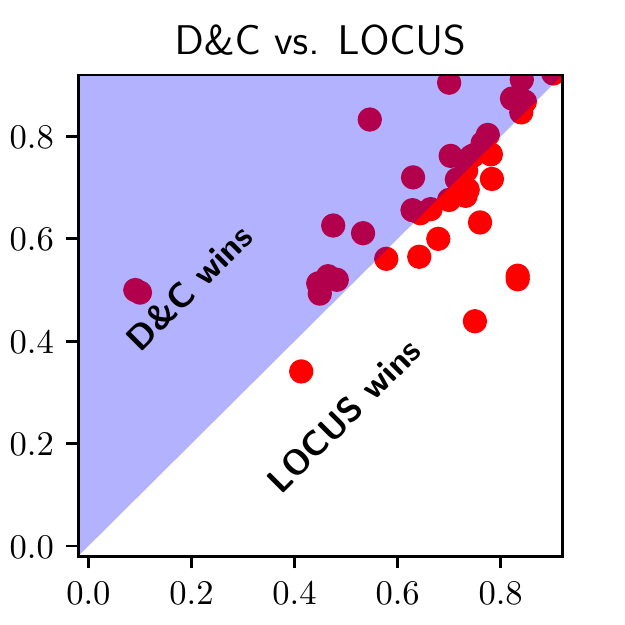}
\endminipage\hfill
\caption{Project-wise performance comparison(X and Y axes show MAP and MRR values, the red dots).}
\label{fig:compareScatter}
\end{figure*}

\subsection{Project-wise performance comparison}
Given the lack of cleaned (i.e., pre-fix activity) data for many projects, our learning merges data from all bug reports for training.
We now investigate the performance on \toolname on localizing bug reports for each project (with training on all project data).

Table~\ref{tab:projectWise} provides details on the performance obtained for each project in our dataset.
It is noteworthy that MAP and MRR performance are significantly varying across projects. While MAP can drop to as low as 0.35 for some projects (e.g., APACHE-WEAVER), it can reach 0.8 for others (e.g., APACHE-LANG). APACHE-IO shows an MRR of 0.91 while JBOSS-JBMETA has an MRR at 0.34. 
The data further shows that this performance is not correlated with the size of the bug reports set of the project.
	 \begin{table}[!h]
\centering
		\scriptsize

\caption{Project wise performance results.}%
\label{tab:projectWise}

\resizebox{\linewidth}{!}{%
\begin{tabular}{l|r|r|r|r|r|r|}
\toprule
				  \multicolumn{4}{r}{Performance} & \multicolumn{3}{|r|}{\% Localized Bug Reports}  \\
				  \midrule
Project & \makecell[c]{\# Bug \\ Reports} & MAP &MRR & Top1 & Top5 &Top10 \\
\midrule
APACHE-CAMEL	&	797	&	0.434	&	0.569	&	44\%	& 72\%	& 80\%\\
APACHE-CODEC	&	28	&	0.79	0&	0.923	&	86\%	 & 100\% &	100\%\\
APACHE-COLLECTIONS	&	26	&	0.776	&	0.874	&	81\% & 92\% &	96\%\\
APACHE-COMPRESS	&	56	&	0.704	&	0.847	&	77\% &	95\% &	100\%\\
APACHE-CONFIGURATION	&	11	&	0.518	&	0.611	&	45\% & 73\% &	82\%\\
APACHE-CRYPTO	&	1	&	1	&	1	&	100\% & 100\% & 100\%\\
APACHE-CSV	    &	8	&	0.66	0   &	0.656	&	38\% &	100\% &	100\%\\
APACHE-HBASE	&	228	&	0.520	&	0.626	&	52\% &	77\%	 & 84\%\\ 
APACHE-HIVE	    &	395	&	0.413	&	0.513	&	39\% & 65\% & 75\%\\
APACHE-IO	    &	43	&	0.878	&	0.911	&	84\% & 100\% & 100\%\\
APACHE-LANG	    &	99	&	0.821	&	0.868	&	84\% & 100\% & 100\%\\
APACHE-MATH	    &	4	&	0.819	&	0.875	&	75\% & 100\% & 100\%\\
APACHE-WEAVER	&	1	&	0.350	&	0.500	&	0\% & 100\% & 100\%\\
JBOSS-ELY	    &	5	&	0.467	&	0.733	&	60\% & 80\% & 100\%\\
JBOSS-ENTESB	&	6	&	0.423	&	0.439	&	17\% & 83\% & 83\%\\
JBOSS-JBMETA	&	10	&	0.298	&	0.341	&	30\% & 30\% & 40\%\\
JBOSS-SWARM	    &	35	&	0.464	&	0.52	0   &	34\% & 77\% & 86\%\\
JBOSS-WFARQ	    &	1	&	1	&	1	&	100\% & 100\% & 100\%\\
JBOSS-WFCORE	&	310	&	0.415	&	0.527	&	41\% & 67\% & 79\%\\
JBOSS-WFLY	    &	461	&	0.413	&	0.493	&	37\% & 62\% & 72\%\\
JBOSS-WFMP	    &	3	&	0.39	0   &	0.528	&	33\% & 100\% & 100\%\\
SPRING-AMQP	    &	67	&	0.561	&	0.720 	&	58\% & 87\% & 93\%\\
SPRING-ANDROID	&	7	&	0.588	&	0.717	&	58\% & 87\% & 93\%\\
SPRING-BATCH	&	239	&	0.543	&	0.684	&	58\% & 87\% & 93\%\\
SPRING-BATCHADM	&	15	&	0.484	&	0.565	&	33\% & 87\% & 93\%\\
SPRING-DATACMNS	&	110	&	0.626	&	0.76	0   &	66\% & 89\% & 94\%\\
SPRING-DATAGRAPH&	2	&	1	&	1	&	100\% & 100\% & 100\%\\
SPRING-DATAJPA	&	103	&	0.586	&	0.762	&	65\% & 92\% & 96\%\\
SPRING-DATAMONGO&	195	&	0.567	&	0.699	&	56\% & 87\% & 94\%\\
SPRING-DATAREDIS&	37	&	0.626	&	0.787	&	68\% & 92\% & 95\%\\
SPRING-DATAREST	&	72	&	0.549	&	0.658	&	53\% & 82\% & 89\%\\
SPRING-LDAP	    &	29	&	0.516	&	0.632	&	45\% & 86\% & 97\%\\
SPRING-MOBILE	&	11	&	0.804	&	0.833	&	73\% & 100\% & 100\%\\
SPRING-ROO	    &	113	&	0.517	&	0.561	&	42\% & 75\% & 84\%\\
SPRING-SEC	    &	242	&	0.604	&	0.676	&	55\% & 83\% & 94\%\\
SPRING-SECOAUTH	&	14	&	0.604	&	0.65	0   &	57\% & 71\% & 79\%\\
SPRING-SGF	    &	19	&	0.615	&	0.716	&	58\% & 95\% & 95\%\\
SPRING-SHDP	    &	14	&	0.672	&	0.763	&	71\% & 86\% & 86\%\\
SPRING-SOCIAL	&	35	&	0.677	&	0.765	&	66\% & 94\% & 94\%\\
SPRING-SOCIALFB	&	11	&	0.68	0   & 	0.803	&	73\% & 100\% & 100\%\\
SPRING-SOCIALLI	&	4	&	0.491	&	0.521	&	25\% & 100\% & 100\%\\
SPRING-SOCIALTW	&	7	&	0.79	0   &	0.905	&	86\% & 100\% & 100\%\\
SPRING-SPR	    &	25	&	0.368	&	0.495	&	36\% & 64\% & 72\%\\
SPRING-SWF	    &	40	&	0.523	&	0.600	&	50\% & 73\% & 78\%\\
SPRING-SWS	    &	66	&	0.57	0   &	0.696	&	59\% & 82\% & 88\%\\
\bottomrule
\end{tabular}
}
\end{table}

We compare the MAP and MRR values per project with that of the state-of-the-art IRBL tools for these projects. Figure~\ref{fig:compareScatter} illustrates how \toolname outperforms every other considered state-of-the-art approach for the large majority of projects.

\subsection{Impact of multi-classification.}

 Our \toolname approach builds and merges the results of multiple classifiers to rank localization files. We investigate the performance of specific classifiers such as the ones built by considering only datasets where a given state-of-the-art tool is exclusively performing well (e.g., {Only-Brtracer}), or the $ INTER$section of datasets where state-of-the-art tools are performing well, or datasets of $UNION$ of bug reports that are localized with Top1 predictions by any state-of-the-art tools, or datasets of $\neg{UNION}$ of bug reports which are localized in Top1 by none of the state-of-the-art tools. We refer to these classifiers as region-specific classifiers. 
Table~\ref{tab:singleVsMultiClassifier} provides MAP and MRR results along with that of \toolname. 
We observe that \toolname outperforms every other region-specific classifiers, followed by UNION. 
This result suggests that the \toolname learning model finds a good way to compute the most effective weights to apply to the similarity scores of IR features. On the other hand, $INTER$ performs lower than $UNION$ perform very well due to the under-representation of a diverse set of bug reports. Finally, we note that region-specific classifiers targeting datasets that are best fitted to a given state-of-the-art tool are not performing well as well. This again suggests that a significant portion of bug reports are not adapted to each such classifiers. Finally, we note that the classifier based on {$\neg{UNION}$} (i.e., no state-of-the-art approach is successful for Top1) can lead to a better performance than some other region-specific classifiers. This finding again confirms that it is indeed necessary to triage the dataset (i.e., dividing based on some rationale).

 \begin{table}[!h]
\centering
		\scriptsize

\caption{Region-classifiers vs Multi-classifier.}%
\label{tab:singleVsMultiClassifier}
\resizebox{1\linewidth}{!}{%
		\begin{tabular}{l|rr|rrrr}

				  \multicolumn{3}{r|}{Performance} & \multicolumn{3}{r}{\% Localized Bug Reports}  \\
				  \midrule
			& MAP & MRR& Top1 & Top5 & Top10\\
        \noalign{\smallskip}\hline\noalign{\smallskip}
\multicolumn{6}{c}{Region-classifiers} \\
\midrule
  OnlyBugLocator & 0.323 & 0.404 & 27\% & 56\% & 67\%\\ 
       OnlyBLUiR & 0.008 & 0.009 & 0.1\% & 1\% & 1\%\\ 
     OnlyAmaLgam & 0.207 & 0.261 & 18\% & 36\% & 44\%\\ 
    OnlyBRTracer & 0.348 & 0.438 & 29\% & 61\% & 73\%\\ 
        OnlyBLIA & 0.367 & 0.454 & 33\% & 61\% & 71\%\\ 
       OnlyLocus & 0.421 & 0.517 & 39\% & 68\% & 77\%\\ 
      BugLocator & 0.483 & 0.595 & 48\% & 74\% & 82\%\\ 
           BLUiR & 0.477 & 0.589 & 47\% & 73\% & 81\%\\ 
         AmaLgam & 0.476 & 0.588 & 47\% & 74\% & 81\%\\ 
            BLIA & 0.480 & 0.593 & 48\% & 73\% & 81\%\\ 
        BRTracer & 0.479 & 0.594 & 48\% & 73\% & 81\%\\ 
           Locus & 0.469 & 0.579 & 46\% & 73\% & 81\%\\ 
           UNION & 0.483 & 0.598 & 48\% & 74\% & 82\%\\ 
     $\neg$UNION & 0.352 & 0.445 & 31\% & 62\% & 73\%\\ 
           INTER & 0.464 & 0.568 & 45\% & 71\% & 79\%\\ 

\midrule
\multicolumn{6}{c}{Multi-classifier}\\
\midrule
           D\&C  & 0.507 & 0.617 & 50\% & 76\% & 84\% \\

		\noalign{\smallskip}\hline

		\end{tabular}

		}
\end{table}

\section{Discussion}
We discuss the insights of our study, the practicality of \toolname and the threats to validity of our results.
\label{sec:discussion}
\subsection{Insights}
{\em On the dividing strategy.} Given the challenge in categorizing bug reports with respect to localization performance, we leveraged state-of-the-art prediction results as proxies to identify groups of bug reports which may share similar properties that are relevant to localization. Concretely, we consider that bug reports, exclusively detected by a given tool (or detected by all tools), share some common characteristics which fit with the different feature sets used by different tools. A potential research direction would consist in further investigating the metrics that could be used to implement other dividing strategies.

{\em On the considered features.} In this work, we focus on common, easily extractable features used by most works in the literature. Nevertheless, we note that there are several recent works which propose other specific features such as code smells~\cite{takahashi2018preliminary} or function call graphs~\cite{wu2014crashlocator}. Although these features have not led to significant improvement of localization performance in one-size-fits-all approaches, they could have a more positive impact in the D\&C approach since region-specific training could properly weight the similarity scores associated to such features for related corner-case bug reports.


\subsection{Practicality}
Validation experiments (see Table~\ref{tab:validationExp}) provide evidence that the \toolname approach is stable: whether training data is small (e.g., 79 bug reports in 2008) or huge (e.g., 4216 bug reports in 2016), the overall performance is stable.
In practice, the training phase which is the most time-consuming can then be done once and be regularly applied to new bug reports.
Because some projects have only one bug report, we have opted to merged all data in a cross-project training scenario. The yielded results are promising and show that \toolname can be leveraged from the start of a development project since training data can be borrowed from other projects. Finally, note that we have performed in-project training as well for big projects such as APACHE-CAMEL: the obtained performance results are similar to when using cross-project training.
		
\subsection{Threats to Validity}
{\bf External validity.} Our study carries a few threats to validity related to the use of Bench4BL where the ground truth of localization may be incomplete (a given bug report may have been fixed by more commits than included in the bench), wrong (some localized files may be wrong as the commit could have been reverted later). The quality of the bug reports may also bias the experiments. Finally, we focus only on Java and the D\&C approach may not generalize to other languages. Nevertheless, these threats are mitigated by the size of the benchmark as well as the inclusion of projects which have been largely investigated in other software mining research works.

{\bf Internal validity.} Our work also carries a number of threats related to the process of cleaning the dataset to consider only post-fix activities. We minimize this threat by using heuristics that are reasonable given the practice of bug reporting in open source communities. Another threat to internal validity is the selection of LightGBM as the core supervised learning algorithm. There is a need to investigate in future work, whether other algorithms will lead to the same conclusions about dividing and conquering with multi-classification. Finally, the presented results are based on merging prediction probabilities without any form of normalisation. Actually, we have used some heuristics to normalize the predictions, but did not notice any change in the performance score.

{\bf Construct validity.} In this study, we hypothesize that the weighting scores of features are the key elements for improving bug localization. However, one threat to validity is that we have leveraged machine learning to estimate these weights: our training step may not actually be modeling the features weights. Finally, we are focusing on comparing the tools, with the threat to validity that the major issue could be rather in the inner IR method.
	
{\bf Conclusion validity.} After dataset curation to remove post-fix activities, the available validation sets contain each a few hundreds or thousands bug reports. Our conclusions are thus threatened. Furthermore, our bug reports may not be heterogeneous indeed. Nevertheless, we minimized these threats by considering the largest dataset ever used in the literature of IRBL.

\section{Related Work}
\label{sec:related}


\subsection{IR methods}
Information Retrieval (IR) generally implies a method used to find data related to a user's needs in various groups of text data. Some of the bug localization studies also use the IR-based technique to find the source code that should be modified using the information from the bug report.
Various approaches are proposed including both simple text models such as Unigram Model (UM)~\cite{song1999general}, Vector Space Model (VSM)~\cite{salton1975vector,wong1985generalized} and sophisticated models such as the Latent Semantic Analysis Model (LSA)~\cite{deerwester1990indexing,dumais2004latent},
the Latent Dirichlet Allocation Model (LDA)~\cite{blei2003latent}, and the Cluster Based Document Model (CBDM)~\cite{xu1999cluster}.
As Rao and Kak~\cite{rao2011retrieval} revealed, simple techniques can provide higher accuracy than sophisticated techniques, leading to a wide use of VSM techniques in recent works.

Thomas et al.~\cite{thomas2013impact} focused on the impact of classifier configurations, studying several parameter values (e.g., code pre-processing, similarity metrics, and term weights) in bug localization task to investigate the performance of the underlying IR methods. Khatiwada et al.~\cite{khatiwada2018just} investigated the performance a new paradigm of information-theoretic IR methods( Pointwise Mutual Information (PMI) and Normalized Google Distance (NGD) )in bug localization tasks.

\subsection{Query Reformulation}
Sisman and Kak~\cite{sisman2013assisting} introduced query reformulation in the context of IR-based bug localization. Chaparro et al.~\cite{chaparro2017using} manually reduce noisy, ineffective queries to reformulated queries that contain only terms that describe observed behaviors, and find that the reformulated queries have much improved performance. Rahman et al. ~\cite{rahman2018improving} incorporate context-aware (i.e., report quality aware) query reformulation into the IR-based bug localization.


\subsection{VSM in IRBL}
Variations of Vector Space Models (VSMs) are used for bug localization as well.
BugLocator~\cite{zhou2012should} uses the revised Vector Space Model (rVSM) to recommend target files to be fixed.
They first made a vector by using keywords extracted from incoming bug report and then compare the vectors to recommend the most probable source code.
The technique computes \textit{SimiScore}, a metric that calculates a similarity between an incoming report and the files fixed by previous bug reports.
Wang et al.~\cite{wang2014compositional} combined a genetic algorithm and VSM to improve the performance of IRBL.
They used Eclipse, SWT, AspectJ and ZXing projects to evaluate their approach.
In the evaluation, this technique achieves 33--48\% accuracy, outperforming previous bug localization approaches.

\subsection{Topic modeling in IRBL}
Topic modeling and semantic analysis are common techniques used in IRBL.
PROMESIR~\cite{poshyvanyk2007feature} utilizes Latent Semantic Analysis (LSI)~\cite{deerwester1990indexing} to identify buggy files.
Lukins et al.~\cite{lukins2010bug} adopted Latent Dirichlet Allocation (LDA)~\cite{blei2003latent} to their approach that models source code topics and showed its effectiveness with a small number of case studies. BugScout~\cite{nguyen2011topic}, on the other hand, builds topic models for both source code and bug reports and compares their distribution to locate files to fix a bug. Takahashi et al.~\cite{takahashi2018preliminary} use code smells to improve bug localization.

\subsection{Stack traces in IRBL}
Stack traces are regarded as a promising information source in bug localization.
Wong et al.~\cite{wong2014boosting} proposed a Brtracer which further considers stack traces in similarity scores.
that separates the source code into segments of a certain size and generates a word vector for each segment, then compare those vectors with the word vector of the target bug report.
in addition to \textit{SimiScore}~\cite{zhou2012should}, this approach leverages the stack trace score that represents the ranking of the top ten file names in a stack trace.
Lobster~\cite{moreno2014use} also uses stack traces to compare with code elements in source code files. CrashLocator~\cite{wu2014crashlocator} focuses more on stack traces together with function call graphs.
to locate a crashing point in a program.
Using the scores, Brtracer recommends files to fix. The evaluation results show that the approach achieves an accuracy of up to 53\%.

\subsection{Feature combinations in IRBL}
Combining existing approaches can improve the performance of IRBL. Amalgam~\cite{wang2014version} uses version history information
building on the intuition that bugs are likely to occur again in files changed more frequently.
BILA~\cite{youm2015bug} takes advantage of source code entities~\cite{saha2013improving}, \textit{SimiScore}~\cite{zhou2012should}, stack trace score~\cite{wong2014boosting}, and version history score.
However, it is difficult to compare whether they have improved on average since there is no experiment on the Eclipse project which is the biggest bug reporting system.
Locus~\cite{wen2016locus}, the most recent technique, proposes fine-grained localization by using commit logs  as well as change hunks in revision history to improve similarity measures.
Then, the technique incorporates token, code entity, and history scores to compute the final similarity score.
The evaluation results suggest that the technique outperforms the existing techniques by 8--10\% and achieves up to 64\% accuracy.
They used SWT, JDT, Tomcat projects to evaluate the approach and obtained around 8\% to 10\% point higher on average and maximum 64\% accuracy. Locus is the state of the art, but since we were not able to use the source code for the experiment, we excluded it from the experimental projects.


\subsection{New approaches to IRBL}
Deep learning techniques also can be leveraged together with IR techniques for bug localization. Lam et al. presented HyLoc~\cite{lam2015combining} and DNNLoc~\cite{lam2017bug}. These approaches use deep neural networks to learn relevancy between tokens in bug reports and code elements in the source code.
In addition, the approaches add an autoencoder to reduce the size of features.
Since the number of tokens in bug reports and source code is often hundreds of thousands, the scalability of neural networks is limited. The autoencoder compresses dimensions of input features.

Other IRBL techniques consider machine learning. Ye et al.~\cite{ye2016mapping} proposed a learning-to-rank approach to bug localization based features representing the degree of suspiciousness. Kim et al.~\cite{kim2013should} dealt with bug report quality to improve bug localization with a two-phase model focusing on high-quality bug reports.

\subsection{IRBL-related studies.}
Closely related to our work, Le et al.~\cite{le2014predicting,le2017will} have proposed a study where they attempt to predict whether the ranked list produced by a bug localization tool is likely to be relevant to the given bug.
They extract various textual and metadata features from 3 old projects and test on two IRBL techniques. They indeed find that it is possible, to some extent, to predict the effectiveness of the considered techniques.
Our work is a generalized and large-scale investigation into the question of IRBL performance.

Saha et al.~\cite{saha2014effectiveness} conducted a study investigating the applicability of an IRBL technique on non-object-oriented code, notably C programs.
They extend a previous approach targeting Java programs to support C code parsing.
They found that IR-based bug localization in C software at the file level is overall as effective as in Java software. They, however, conclude that  using program structure information to tune localization is less relevant to C software than for Java software.

Wang et al.~\cite{wang2015evaluating} have conducted
an analytical study and a user study on IRBL techniques
to assess their usefulness. Focusing on a single technique, BugLocator, and
four common projects from previous studies, they report that the information
needed for IR-based techniques to be effective is often not
available in bug reports.
Their user study further suggests that
even when high-quality bug reports
are available and IR-based techniques can ``perfectly'' rank bug locations, they may still benefit developers
only marginally since high-quality
bug reports are often good enough to guide developers to
the file, which can be located without any additional
help.
They also discuss that suspicious file ranking by IRBL techniques
may not help speed up the localization
of the bug within that file, which could be the
most time-consuming part of debugging.
Nevertheless, as in many software engineering-related tasks, automated IRBL can accelerate the realization of other endeavors (e.g., improve the scalability of automated program repair).

Recently, Lee et al~\cite{lee2018bench4bl} have proposed an extensive benchmark for IRBL. They used this benchmark to offer a clear view on the performance of state-of-the-art working tools. In our study, we leveraged their dataset and further curated it to remove any post-fix activities data.

\section{Conclusions}
\label{sec:conclusion}
We have proposed \toolname, a novel IRBL approach which adaptively learns to compute the weight to associate to similarity scores of IRBL features. To that end, we leverage a gradient boosting supervised learning technique to build multi-classifiers by training on homogeneous subsets of bug localization datasets. 
In practice, we have performed a large scale empirical study which revealed that state-of-the-art tools, which mainly differ by the features that are considered, appear to be fit for specific bug reports. Thus, we leverage the assessment results of six state-of-the-art tools as a metric for splitting the dataset and allowing a meaningful training of specialized classifiers whose outputs are then combined to produce an accurate ranking of localization recommendations. 
Comparing to state-of-the-art tools, \toolname shows higher performance on Bench4BL, currently the most comprehensive bug localization dataset in the literature.
Typically, our validation experiments yield an MAP score of 0.52, and an MRR score of 0.63 with a curated version of Bench4BL. Comparison against the state-of-the-art shows that \toolname provides a substantial performance improvement of MAP and MRR over all tools: MAP is improved by between 4 and up to 10 percentage points, while MRR is improved by between 1 and up to 12. Finally, we note that \toolname is stable in its localization performance: around 50\% of bugs can be located at Top1, 77\% at Top5 and 85\% at Top10.

Future along this direction could consider including more classifiers trained on corner-case bug reports which can be discovered by tools  which include features such as crashes or code smells. 
Similar experiments can be performed at the method level to assess the performance on finer-grained bug localization as attempted by Locus, although with poor performance. 
Finally, the research community can benefit from a reverse engineering of the exclusive successful localization results by various state-of-the-art to formally model the characteristics of the associated bug reports, to improve other research lines, notably on duplicate bug detection, bug triaging, etc.
\paragraph*{Availability} The codebase of \toolname, all pre-computed feature matrices, as well as produced classifier prediction probabilities are available at:  
\begin{center}
	\url{https://github.com/d-and-c/d-and-c}.
\end{center}

\balance
\bibliographystyle{spmpsci}
\bibliography{bib/ref}

\end{document}